\documentclass[10pt, conference,usletter]{IEEEtran}
\usepackage{cite}
\ifCLASSINFOpdf
\else
\fi

\usepackage{graphicx}
\usepackage{epstopdf}
\usepackage{epsfig}
\usepackage[cmex10]{amsmath}
\usepackage{amssymb}
\usepackage{amsthm}

\usepackage{balance}
\usepackage{flushend}

\usepackage{algorithm, algorithmic}
\makeatletter
\newcommand\fs@norules{\def\@fs@cfont{\bfseries}\let\@fs@capt\floatc@ruled
  \def\@fs@pre{}%
  \def\@fs@post{}%
  \def\@fs@mid{\kern3pt}%
  \let\@fs@iftopcapt\iftrue}
\makeatother
\restylefloat{algorithm}

\usepackage[font=footnotesize, belowskip=2pt, aboveskip=0pt]{caption}
\usepackage[square, comma, numbers]{natbib}
\usepackage{algorithm}
\usepackage{algorithmic}
\usepackage{subfigure}

\usepackage{array}
\ifCLASSOPTIONcompsoc
  \usepackage[caption=false,font=normalsize,labelfont=sf,textfont=sf]{subfig}
\else
  \usepackage[caption=false,font=footnotesize]{subfig}
\fi

\usepackage[draft,footnote,nomargin]{fixme}
\usepackage{url}
\begin{document}
\title{Backward-Shifted Coding (BSC) based on Scalable Video Coding for HAS}
\author{\IEEEauthorblockN{Zakaria Ye\IEEEauthorrefmark{1},
Rachid El-Azouzi\IEEEauthorrefmark{1},
Tania Jimenez\IEEEauthorrefmark{1} and
Francesco De Pellegrini\IEEEauthorrefmark{2}
} 
\IEEEauthorblockA{\IEEEauthorrefmark{1}LIA/CERI, University of Avignon, 84000, Avignon, France, 
\IEEEauthorrefmark{2}CREATE-NET, 30123, Trento, Italy \\
Email: \{zakaria.ye, rachid.elazouzi, tania.jimenez\}@univ-avignon.fr} fdepellegrini@create-net.org}
\maketitle
\thispagestyle{plain}
\pagestyle{plain}
\begin{abstract}
The main task of HTTP Adaptive Streaming is to adapt video quality dynamically under variable network conditions. This is a key feature for multimedia delivery especially when quality of service cannot be 
granted network-wide and, e.g., throughput may suffer short term fluctuations. 

Hence, robust bitrate adaptation schemes become crucial in order to improve video quality. The objective, in this context, is to control the filling level of the playback buffer and maximize the quality of the video, while avoiding unnecessary video quality variations.

In this paper we study bitrate adaptation algorithms based on Backward-Shifted Coding (BSC), a scalable video coding scheme able to greatly improve video quality. We design bitrate adaptation algorithms that balance  video rate smoothness and high network capacity utilization, leveraging both on throughput-based and buffer-based adaptation mechanisms. 

Extensive simulations using synthetic and real-world video traffic traces show that the proposed scheme  performs remarkably well even under challenging network conditions.
\end{abstract}

\section{Introduction}
In the last years, smartphones and other mobile devices have emerged as one dominant technology for daily access to Internet services. This, combined with the ever increasing  broadband access supplied by operators has triggered pervasive demand on video streaming mobile services. In turn, this requires the exploration of novel approaches on video content delivery. To afford video streaming services at sustainable costs, the idea of adjusting the bit rate of video traffic depending on the (time-varying) available bandwidth has been actively investigated during the recent years. This technique is commonly referred to adaptive streaming technology. At the industrial level, many adaptive video streaming solutions exist. They are now undergoing a standardization process under the Dynamic Adaptive Streaming over HTTP (DASH) initiative. DASH will include
existing solutions such as Microsoft's smooth streaming, Adobe's HTTP dynamic streaming and Apple's live streaming \cite{team2002coding}. In order to fully exploit the 
potential of DASH, though, new challenges arise for content providers, operators and device manufacturers. One of such challenges is the need to accurately assess users' Quality of Experience (QoE) in order to enhance service provisioning and optimize adaptation to network conditions.

Actually, the key concept in DASH is to dynamically adapt the video quality to the network bandwidth. This is done in order to cope with multiple playback interruptions. Those are likely to occur when the video quality is kept the same during the whole video session irrespective of possibly highly variable network conditions, e.g., those typical of mobile wireless connections. In DASH, a single video file is divided into smaller chunks of fixed playback duration called segments. Each segment is encoded at various bitrate levels (called representations). This is done using a specific compression algorithm or codec (e.g., H264/AVC). Then, given the available network bandwidth, a segment is selected with the appropriate bitrate. With the Scalable Video Coding (SVC) compression algorithm (extension of Advanced Video Coding), the video source is encoded in one base layer (BL) and one or more optional enhancement layers (ELs), as depicted in Fig.~\ref{segcoding}. The base layer is always provided. Then, given the available network bandwidth, the client adaptation engine adds the appropriate number of enhancement layers in order to improve the video quality/SNR, resolution and frame rate. 

The design  goal of DASH is to simultaneously obtain high performance over different key metrics including buffering delay, playback interruptions, average bitrate (video quality) and temporal variability of streaming quality. However, in an environment subject to highly variable throughput, attaining high performance across all these metrics
is still considered a great challenge. In this paper, we propose a novel bitrate adaptation scheme which is based on our Backward Shifted Coding (BSC) introduced in \cite{BSCcoding}. This BSC system makes HTTP Adaptive Streaming (HAS) more robust to rapid fluctuations of the network capacity and provides more flexibility in increasing the quality of video without playback interruptions. The basic idea of BSC is to shift the base layer and its enhancement layers so that when an interruption of playback buffer occurs, the base layer can still be played. 

In \cite{BSCcoding}, we have characterized the performance of BSC with a single video quality, i.e., in the case when the quality of the segments does not change. In that case we have showed that BSC can improve the video quality and reduce the probability of  playback interruptions. However, the performance of the BSC scheme has been assessed without exploiting on the full potential of DASH, which is capable to tune layer quality at runtime. 

In this paper, we incorporate a new version of BSC in HTTP adaptive streaming. Doing so we are able to strike the balance between responsiveness and smoothness in DASH. More in detail, this new version of BSC contains two layers: the {\em low layer segment}, which delivers only the base layer or the base layer with a minimal set of enhancement layers, and the {\em top layer segment} that contains only enhancement layers. During the video transmission, the two segments are shifted in time. Hence, our main focus in the following is the adaptation problem, i.e., how to jointly match the video quality of each layer (low layer and top layer) of the two shifted segments to the network conditions. Our proposed adaptation methods select the appropriate bitrates for both segments by adding the appropriate number of enhancement layers. Through extensive simulations we show that this BSC system performs remarkably well even under high throughput variability. This is due to the key property of this novel scheme. In fact, the DASH protocol can leverage on the time difference of the two BSC layers, which increases diversity. In turn, this mitigates the impact of inaccurate capacity estimation on HAS.  In summary, this paper makes the following key contributions
\begin{itemize}
\item We adapt BSC for HTTP Adaptive Streaming to provide a practical solution able to improve the QoE delivered by the current DASH;
\item  We propose a novel rate adaption algorithm that balances the need for maximum quality, video rate smoothness  while avoiding the risk of stalls; 
\item  We detail throughput-based and buffer-based adaptation schemes and verify the improvement of video quality even under high throughput variability;   
\item We show that BSC brings more robustness for buffer based schemes in realistic network environment.
\end{itemize}

The outline of the paper is as follows. In Section \ref{model} we describe the Backward Shifted Coding system and its mapping to the DASH system. Section \ref{bsc_sys} details the bitrate adaptation in Backward Shifted Coding including the pseudo-codes of our proposed adaptation algorithms. Section \ref{num} presents the simulations framework and the numerical results. Section \ref{relatedsec} gives an overview of the existing bitrate adaptation algorithms used in DASH literature. Finally, Section \ref{conclusion} concludes the paper.

 \section{Backward-Shifted Coding}
\label{model}
\begin{figure}[t]
\begin{center}
\includegraphics[scale=0.55]{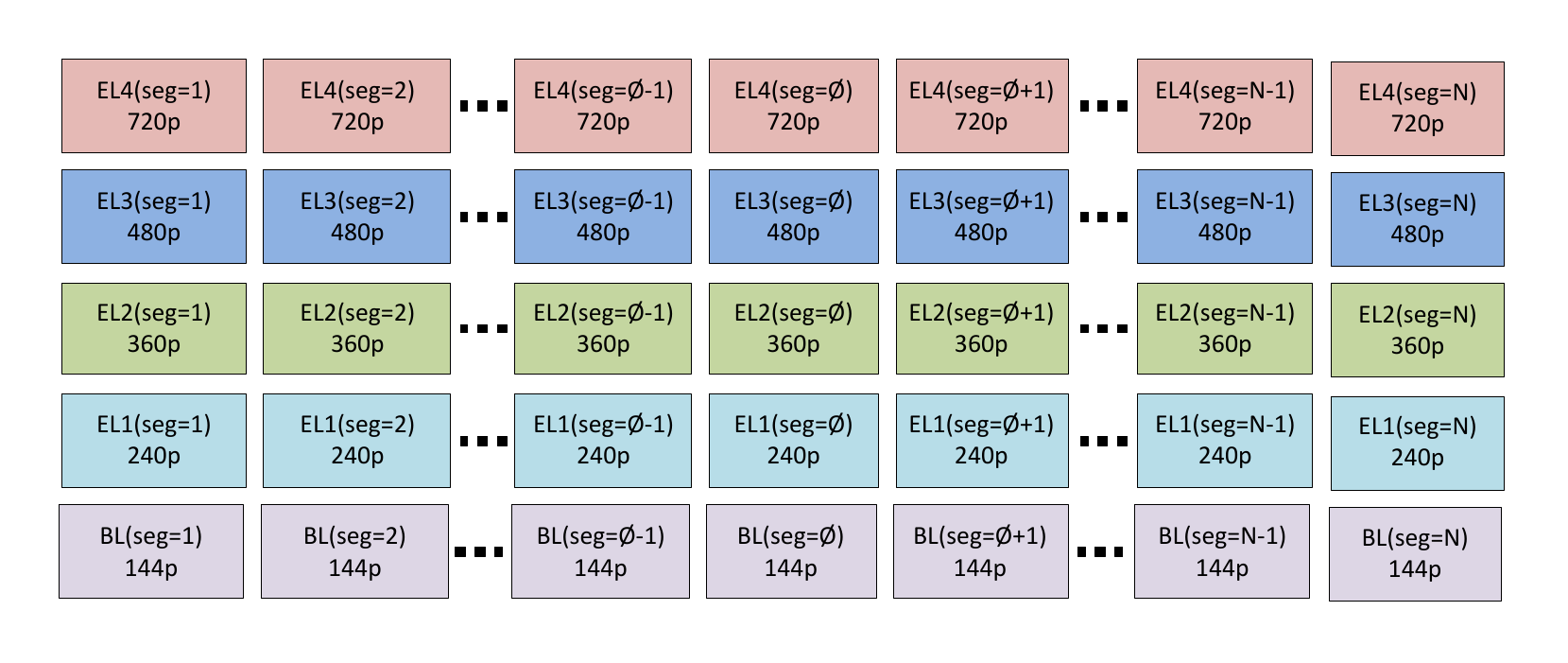}
\caption{Segments encoding with SVC}
\label{segcoding}
\end{center}
\end{figure}

\begin{figure*}[hbt!]
\begin{center}
\includegraphics[scale=0.8]{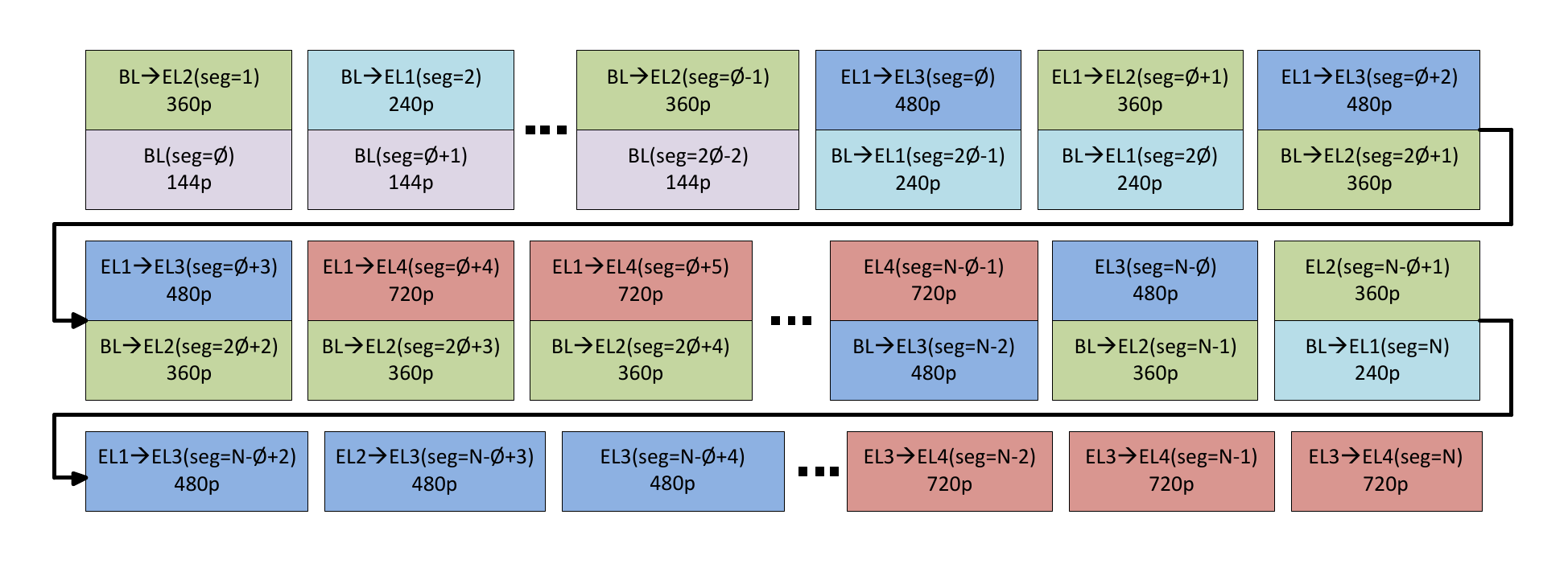}
\caption{Segments transmission with Backward-Shifted Coding: the low layer segments contain the base layer (and possibly some enhanced layers) and are transmitted before the corresponding top layer segments, which follow after $\phi-1$ blocks; the initial $\phi$ blocks carry only lower layer segments; the notation $BL \rightarrow EL_j$ indicates all segments  $BL,EL_1,EL_2,...,EL_j$ and $EL_i \rightarrow EL_j$ indicates $EL_i,EL_{i+1},...,EL_j$}
\label{SVC_BSC}
\end{center}
\end{figure*}

Hereafter we shall provide an overview of BSC and we shall briefly describe our DASH-compliant extension.
The BSC scheme is fully client driven. The main idea of the scheme is to send a complete segment 
(base layer and possibly some enhancement layers) together with the enhancement layers of another segment. 
We call the first one, \textit{lower layer segment} and the second one is the \textit{top layer segment}. 
During the video transmission, the two segments are shifted  in time by a constant offset. We denote
 $\phi$ the offset between the two segments. Thus, each segment $k$ has its enhancement layers in segment 
 $k+\phi-1$ (Fig.~\ref{SVC_BSC}). We call {\em block} $k$ the combination of segment $k+\phi-1$ (lower layer) 
 and of enhancement layers of segment $k$ (top layer). Therefore, should the enhancement layer be missed, the player 
 can still playout the lower layer segment which is sent in advance with low quality. 

The advantage of the BSC scheme is apparent if we consider the decoding operations at the user side, i.e., when incoming bits are reassembled into video frames by the decoder. The advantage compared to the basic SVC scheme in Fig.~\ref{segcoding} is that in plain SVC, when lower layer segment $k$ is transmitted, it is decoded to render the segment with a given quality. Later, if other enhancement layers of this segment are received, the segment is decoded again to increase its quality. BSC does not need to perform repeated decoding  since each block is received only once, i.e., base layer and related enhancements layers.

The BSC scheme can be naturally adapted to DASH: under DASH/SVC, video servers store each tagged video into  segments. For multi-layer codecs, such segments consists of a base layer and multiple enhancement layers. BSC requires to compound layers and to defer the transmission of top layer segments. Conversely, bitrate adaptation algorithms have not been standardized yet in DASH. The aim is to choose a bitrate ensuring good video quality and prevent video playback interruptions. They fall into two categories: the {\em throughput-based} approaches and the {\em buffer-based} approaches. Some schemes \cite{thang2014evaluation, yin2014toward} may actually fall in both categories since they leverage on the estimation of the network throughput in combination with buffer-based mechanisms. 

 The main idea behind throughput-based schemes is that the MPEG-DASH client performs an estimation of the available bandwidth for the requested segments \cite{isoiecsegment, stockhammer2011dynamic}. Then, based on the network throughput and the playout buffer occupancy level, an adaptation engine chooses the highest possible bitrate compatible with the available throughput in order to avoid possible playback interruptions. The simplest way to estimate the available throughput is to compute the segment throughput after it is completely downloaded. This is a standard throughput measure called {\em instant throughput} \cite{thang2014evaluation}. This method is simple and fast to  react to the throughput variations but not accurate. Conversely, buffer-based methods leverage on the size of the buffer, with the aim of keeping it at a given nominal level.
 
In this context, the adaptation engine for BSC may request the two segments at different bitrates, i.e, one  bitrate for the low layer segment (base and enhancement layers) and one bitrate for the top layer segment which contains only enhancement layers.

\section{BSC with Bitrate Adaptation}
\label{bsc_sys}

\subsection{System Description}
\label{system}

\begin{figure}[b!]
\begin{center}
\includegraphics[scale=1]{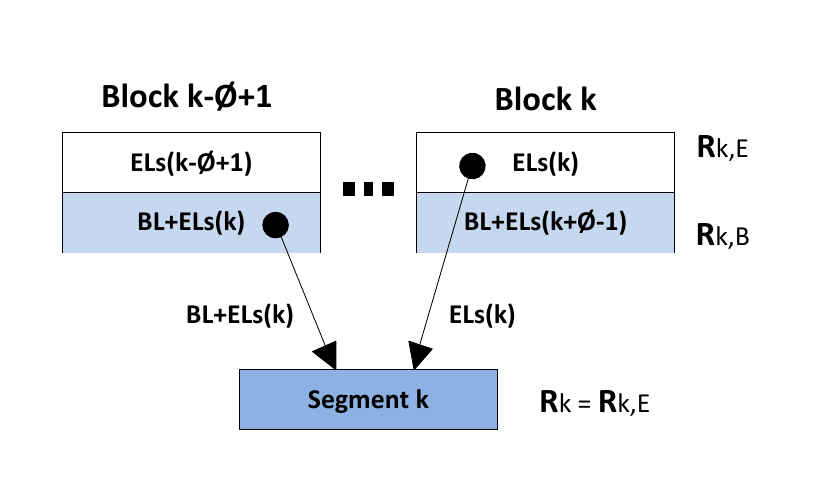}
\caption{Decoding of segment $k$: uses lower layer segment of block $k-\phi+1$ (containing base layer and possibly some enhancement layers) and top layer segment of block $k$ (containing enhancement layers only).}
\label{blockToSegment}
\end{center}
\end{figure}

\begin{figure*}[hbt!]
\begin{center}
\includegraphics[scale=0.8]{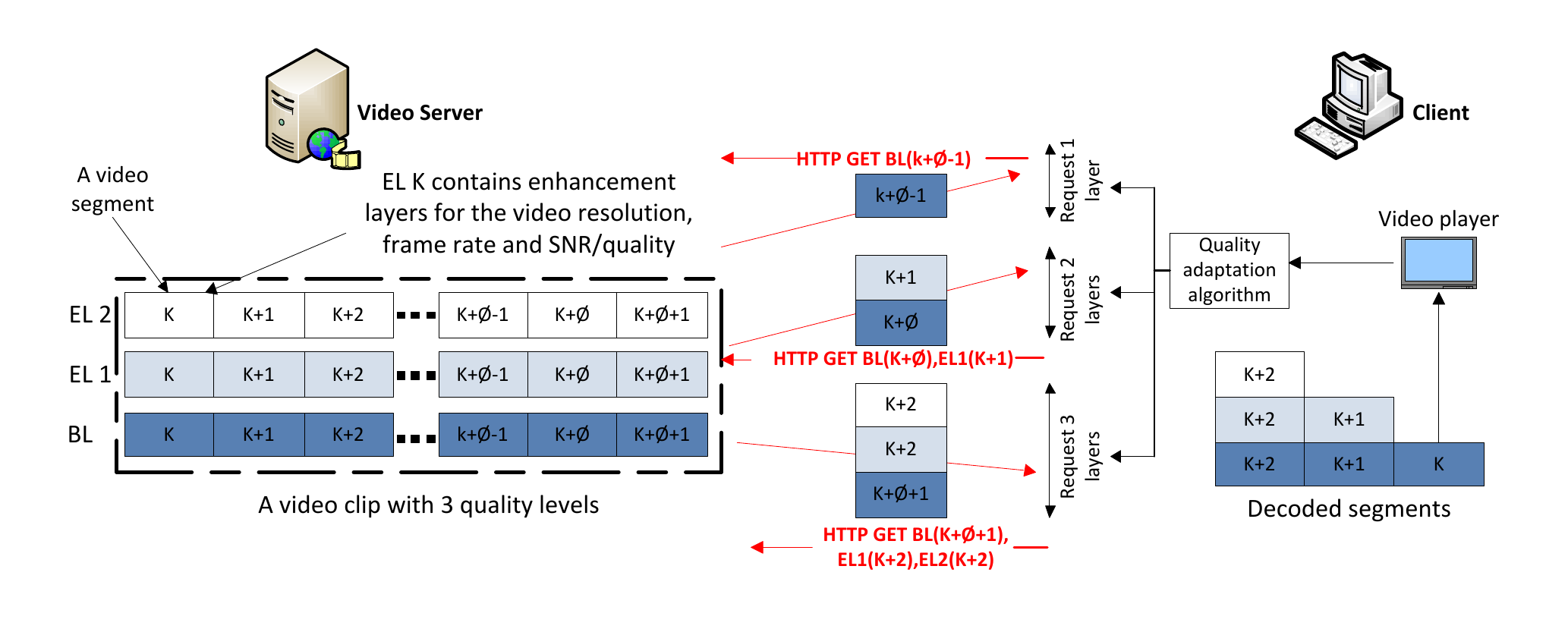}
\caption{Principle of adaptive HTTP streaming with DASH/BSC}
\label{systemfig}
\end{center}
\end{figure*}
In this section, we develop a video rate adaption algorithm suitable for the Backward-Shifted Coding described in section~\ref{model}. The server holds the 
Media Presentation Description (MPD), the media segments and it hosts a HTTP server. In the Backward-Shifted Coding, the media segments are encoded using 
H264/SVC. As shown in Fig. \ref{blockToSegment}, block $k$ contains segment $k+\phi-1$ (lower layer segment) and enhancement layers of segment $k$ (top layer segment). 
The information related to the media segments, e.g., the available video bitrates, are described in the MPD file. The offset $\phi$ has also to be added to the MPD.

Each time a user requests the video, a HTTP connection is established with the server. The MPD file is sent first before any video data is transmitted. The video blocks are downloaded into a playback buffer, which contains downloaded segments but are  not yet displayed by the playout (Fig. \ref{systemfig}). As shown in Fig. \ref{blockToSegment},  after  block  $k$ is  downloaded, segment $k$ can be decoded  using the lower layer segment from block $k-\phi+1$ and the enhancement layers from block $k$. Please observe that the index of block $k$ refers to the index of the segment to which the upper layer belongs to.  
 
Let $N$ be the number of segments contained in the video file. Each segment contains $L$ seconds of 
video and it is encoded  at different bitrates.  

In standard SVC playout, a set of available bitrate levels per segment $\mathcal{R}$ 
corresponds to selecting the base layer and a certain number of enhancement layers. In the BSC system, 
the playout downloads the BSC block $k$ with the bitrates $(R_{k,E}, R_{k,B}) \in \mathcal{R}^2$. In particular we denote:  
\begin{itemize}
\item $R_{k,E}$ is the bitrate of segment $k$  by {\em including the lower layer segment}, which 
is received through block  $k-\phi+1$
\item $R_{k,B}$ is the bitrate of the lower layer segment $k+\phi-1$ 
(which contains base layer and some enhancement layers). 
\end{itemize}
Note that, with this notation, when we refer to the condition $R_{k,E} = R_{k-\phi+1,B}$, we mean that 
no enhancement layers are transmitted in block $k$.

Let $d_{k,E}$ and $d_{k,B}$ be, respectively, the size  of the enhancement layers segment and the size 
of the lower layer segment in block $k$. Thus 
$$
d_{k, E} =  (R_{k,E}-R_{k-\phi+1,B}) L, \mbox{ and }d_{k, B} = R_{k,B} L 
$$ 
so that the corresponding rate for block $k$ given by $R_k=R_{k,E}-R_{k-\phi+1,B}+R_{k,B}$. Clear, the set of all possible block bitrates is still $\mathcal{R}$.

\subsection{Adaptation methods in BSC}
\label{algos}

The goal of the bitrate adaptation is to maximize the quality of experience of the video streaming user 
depending on four key parameters: the startup delay, the playback interruption, the mean video bitrate 
and the bitrate switching. 

\begin{figure}[hbt!]
\begin{center}
\includegraphics[scale=0.6]{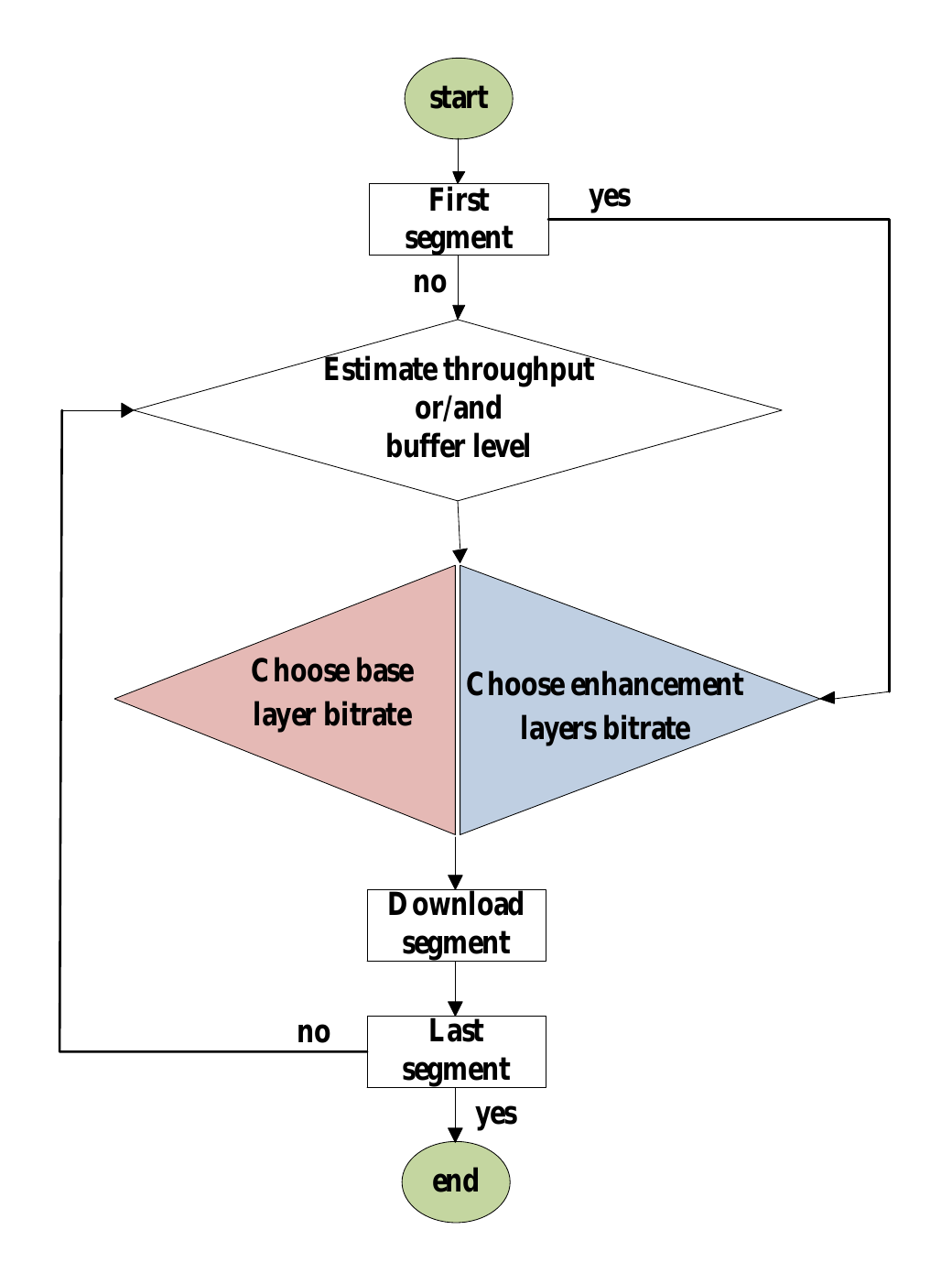}
\caption{Flow diagram for throughput-based and buffer-based adaptation algorithms for BSC.}
\label{Badap}
\end{center}
\end{figure}
We propose bitrate adaptation methods to choose the suitable bitrates for block $k$ (Fig. \ref{Badap}). 
We denote  by $R_{min}$ and $R_{max}$  the smallest and the highest bitrate respectively in the set of available bitrates $\mathcal{R}$. We let $\hat{A}_t$ and $\mathcal{B}_k$  be, respectively,  the estimated throughput after the download of the segment $k-1$ and the current playout buffer occupancy measured in seconds of video content.  
The estimated throughput can be generated based on several techniques such as exponential average or weighted average in order to mitigate short-term fluctuations caused by the lower layers. 

In order to select bitrates $R_{k,B}$ and $R_{k,E}$, we are inspired from the two approaches described in Sec.~\ref{model}, namely the buffer-based and the throughput-based approach in order to evaluate the performance 
of the BSC scheme. This results into two algorithms: the throughput-based BSC algorithm (TB-BSC) and the buffer-based BSC algorithm (BB-TSC).

\textbf{The throughput based approach}. We distinguish two cases based on the block index: $k < \phi$ and $k \geq \phi$.

\begin{figure}[b!]
\begin{center}
\includegraphics[scale=0.75]{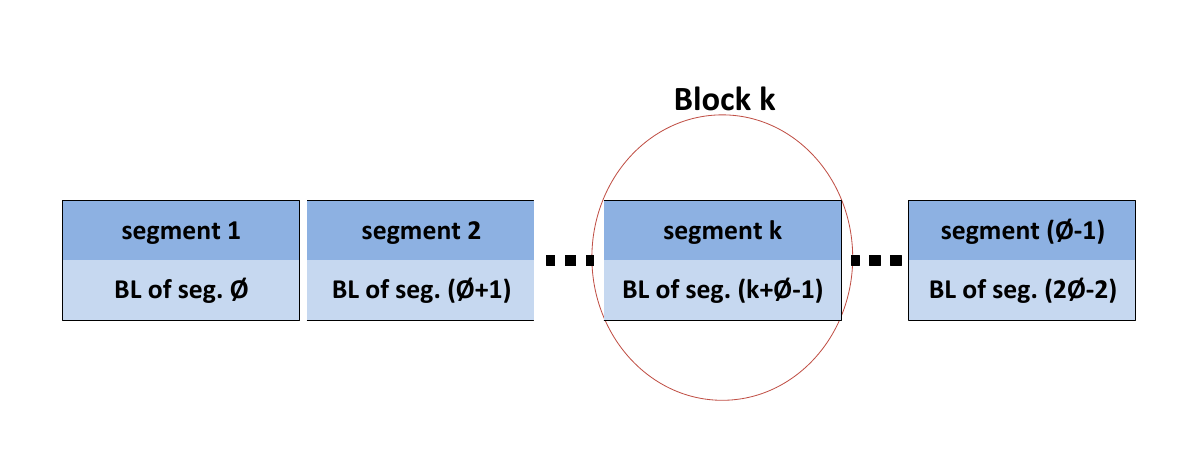}
\caption{Block $k$ contains segment $k$ and BL of segment $k+\phi-1$ for $k < \phi$}
\label{fisrt_seg}
\end{center}
\end{figure}

{\noindent \bf  Case $0\leq k \leq \phi - 1 $.} For the $\phi-1$ first blocks, see Fig.~\ref{fisrt_seg}, each block contains  1) the whole (lower layer) segment $k$ and 2) the lower layer segment $k+\phi-1$ but at minimum bitrate $R_{k,B}=R_{min}$. Thus, for the first $\phi-1$ blocks, the bitrate adaptation concerns only the whole segment $k$ and must be operated such in a way that $R_k + R_{min} \leq \hat{A}_t$ where $R_k$ is the bitrate of the whole segment $k$.

By assigning a minimum bitrate, $R_{min}$, to the lower layer segment $k+\phi-1$, the startup delay is not greatly affected by the BSC scheme. Doing so, we immediately maximize the bitrate of the segments $1\leq k \leq \phi-1$ -- for which no enhancement layers are expected later on -- and we defer the bitrate enhancement of the lower layer segments $\phi \leq k \leq 2 \phi-2$ using the upper layer segment carried by block $k+\phi-1$.

{\noindent \bf  Case $k \geq \phi$.} The pseudo-code for this part of the TB-BSC adaptation algorithm is provided in Alg.~\ref{alg1}. It is interesting to observe that, in our TB-BSC scheme, we shall also leverage on information on the buffer level occupancy. 

We assume that it is invoked repeatedly each time $t$ a block is downloaded; it starts immediately after the download of BSC block $k-1$ is completed. 

Let $\phi_{t}=\phi \cdot L$: it represents the offset in seconds between the lower layer segment and its enhancement layers. When the buffer size (in seconds) is not larger than $\phi_{t}$ (Line $3$), we no longer need to send the enhancement layers segments because their corresponding segments are already been played by the playout. In that case, the bitrate selection is equivalent to DASH/SVC (Line $5$ to $14$). When 
$\mathcal{B}_k > \phi_{t}$ (Line $15$ on), the adaptation is done on both the lower layer segments and the enhancement layers segments. 

In the worst case, i.e. Line $15$, when the estimated throughput is lower than $R_{min}$, the selected bitrate for the lower layer segment in block $k$ is $R_{min}$ and no enhancement layers are sent, i.e., $R_{k,E}=R_{k-\phi+1,B}$. 

We denote by $R_{t-}$, the highest available bitrate compatible with the estimated throughput. In the same way, $R_{t+}$ is the smallest available bitrate regarding the estimated throughput.
\[
\left\{
\begin{array}{l c l}
R_{t-} &=& \{\max \{ R_i \} : R_i \leq \hat{A}_t\} \\
R_{t+} &=& \{\min \{ R_i \} : R_i \geq \hat{A}_t\}
\end{array}
\right.
\]

When the estimated throughput is lower than the bitrate of the lower layer segment in the previous block $k-1$, the bitrate of the lower layer segment in the next block $k$ is set to $R_{t-}$. And the bitrate of the enhancement layers segment in the next block $k$ is the maximum between $R_{t+}$ and $R_{k-\phi+1,B}$ (Line $19$ and $20$, respectively). It is worth remarking that in this case, the selected bitrate for the lower layer segment is not larger than the estimated throughput in order to prevent playback interruptions. But, we observe that the bitrate of the enhancement layers segment is larger than the estimated throughput. Indeed, we do not risk playback interruptions here: in fact the buffer level is large enough ($\mathcal{B}_k > \phi_{t}$). 

\begin{algorithm}[H]
\begin{footnotesize}
 \caption{TB-BSC Algorithm for $k \geq \phi$} \label{alg1}
 \begin{algorithmic}[1]
 \renewcommand{\algorithmicrequire}{\textbf{Input:}}
 \renewcommand{\algorithmicensure}{\textbf{Output:}}
 \REQUIRE $\hat{A}_t$: Estimated throughput of block $k-1$ \\
 		  $R_{k-1,B}$: Bitrate of lower layer segment in block $k-1$ \\
 		  $R_{k-\phi+1,B}$: Bitrate of lower layer segment in block $k-\phi+1$ \\
 		  $\mathcal{B}_k$: Buffer occupancy in seconds
 \ENSURE  $R_{k,B}$: Bitrate of lower layer segment in block $k$ \\
 		  $R_{k,E}$: Bitrate of enhancement layer in block $k$
  \STATE $R_{t-}\gets \{max(R_i) : R_i \leq \hat{A}_t\}$
  \STATE $R_{t+}\gets \{min(R_i) : R_i \geq \hat{A}_t\}$
  \IF {$\mathcal{B}_k \leq \phi_{t}$}
  	\STATE $R_{k,E} := R_{k-\phi+1,B}$ //no enhancement layers
  	\IF {$\hat{A}_t \leq R_{min}$}
  		\STATE $R_{k,B} := R_{min}$
  	\ELSIF {$\hat{A}_t < R_{k-1,B}$}
  		\STATE $R_{k,B} := R_{t-}$ 
  	\ELSIF {$R_{k-1,B} < R_{max}$}
  		\STATE $R_{k,B} := R_{k-1,B}^{\uparrow}$
  	\ELSE
  		\STATE $R_{k,B} := R_{max}$
  	\ENDIF
  \ELSE
  	\IF {$\hat{A}_t \leq R_{min}$}
  		\STATE $R_{k,B} := R_{min}$
		\STATE $R_{k,E} := R_{k-\phi+1,B}$
  	\ELSIF {$\hat{A}_t < R_{k-1,B}$}
  		\STATE $R_{k,B} := R_{t-}$ 
		\STATE $R_{k,E} := max(R_{k-\phi+1,B}, R_{t+})$ 
  	\ELSIF {$R_{k-1,B} < R_{max}$}
  		\STATE $R_{k,B} := R_{k-1,B}^{\uparrow}$
		\STATE $R_{k,E} := max(R_{k-\phi+1,B}, R_{k-1,E}^{\uparrow})$
  	\ELSE
  		\STATE $R_{k,B} := R_{max}$
		\STATE $R_{k,E} := R_{max}$
  	\ENDIF
  \ENDIF
 \end{algorithmic}
 \end{footnotesize}
\end{algorithm}

When the available throughput increases compared the previous block (Line $21$), we increase the bitrate in a smooth manner in order to avoid sudden video quality transitions \cite{akhshabi2011experimental}. In practice, when the estimated throughput is higher than the bitrate of the lower layer segment in the block $k-1$, the selected bitrate of the lower layer segment in the block $k$ is increased to a higher bitrate, i.e., $R_{k,B}=R_{k-1,B}^{\uparrow}$, (Line $22$). The bitrate of the enhancement layers of block $k$ is increased to a higher bitrate as well (Line $23$).
\begin{figure}[t!]
\begin{center}
\includegraphics[scale=0.7]{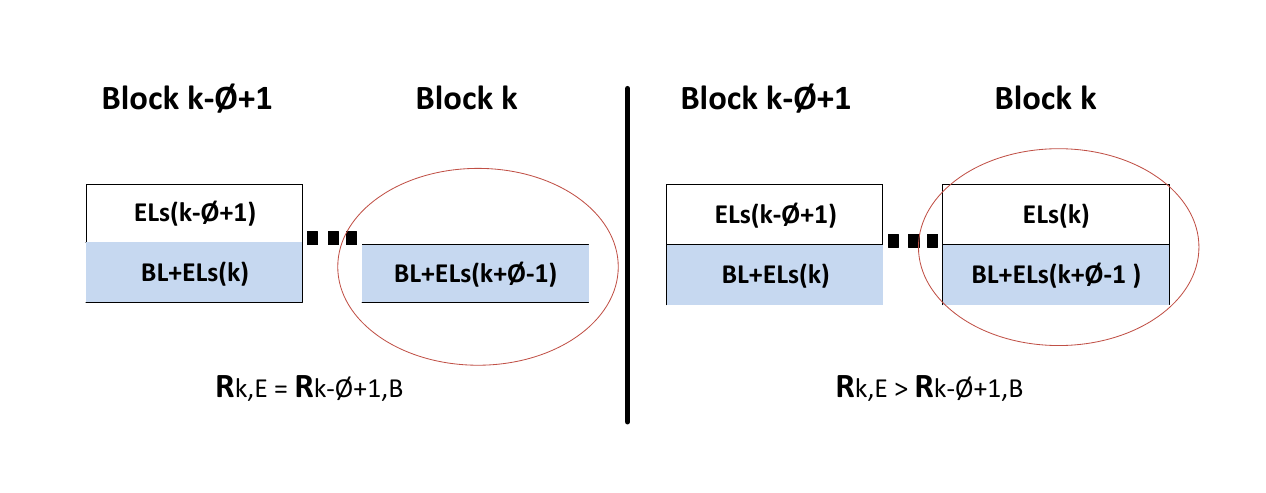}
\caption{If $R_{k,E} = R_{k-\phi+1,B}$, no enhancement layers are transmitted in block $k$, otherwise, the necessary number of enhancement layers are added to block $k$ to reach bitrate $R_{k,E}$}
\label{enhan_fig}
\end{center}
\end{figure}
Note that, when $R_{k,E} > R_{k-\phi+1,B}$, the necessary number of enhancement layers are added to the segment in block $k$ to reach the bitrate $R_{k,E}$. This is illustrated in Fig. \ref{enhan_fig}.  

Finally, in the Alg.\ref{alg1}, the bitrate of the lower layer segment in block $k$ is $R_{t-}$ and the bitrate of the enhancement layers segment is $R_{t+}$. However, when $R_{k,B}=R_{t-}$ and $R_{k,E}=R_{t-}$, the Backward-Shifted Coding system is equivalent to the DASH/SVC system. 

\textbf{The buffer based approach}: 
the use of buffer occupancy to select the segments' bitrate is a technique used by several schemes in the literature \cite{miller2012adaptation, muller2012evaluation, akhshabi2012experimental}. Typically, buffer thresholds are set (either two or three thesholds) and decisions on the bitrate are taken according to the level of current buffer occupancy with respect to such thresholds. Some of these methods use also the estimated throughput to smooth bitrate variations. Let us call BBA-0 this group of bitrate adaptation methods. 

However, there exist another group of buffer-based algorithms where an adjustment function is used to pick the appropriate bitrate \cite{huang2014buffer, huang2014bufferphd}. Let us call them BBA-1: compared to BBA-0, 
they do not perform throughput estimation, thus avoiding the related estimation errors. This method for 
bitrate selection is the basis of our BB-BSC algorithm. We describe first the application to BSC of the template algorithm introduced in \cite{huang2014buffer}, shortly BBA-1. Then we specialize it to match the specific features of BSC and derive BB-BSC. BB-BSC will be finally composed of two procedures, one for the lower layer segments and one for the top layer segments. Those are reported in Alg.~\ref{alg2} and Alg.~\ref{alg3}, respectively.

We have two buffer thresholds $r$ and $c$ where $r$ is the reservoir and $c$ is the cushion in seconds of video content. The bitrate selection is based on an adjustment function $F$ where $F(\mathcal{B}_k)=R_{min}$ for $\mathcal{B}_k \leq r$ and $F(\mathcal{B}_k)=R_{max}$ for $\mathcal{B}_k \geq r+c$. Then, given the current buffer occupancy $\mathcal{B}_k$, $F(\mathcal{B}_k)$ is computed to select the bitrate of the next segment. We use the following function $F$ as in \cite{yin2014toward}:
\[
F(\mathcal{B}_k) =
\left \{
\begin{array}{l l l}
R_{min} & \quad \mathcal{B}_k \leq r \\
R_{max} & \quad \mathcal{B}_k \geq r+c \\
R_{min}+\frac{\mathcal{B}_k-r}{c}(R_{max}-R_{min}) & \quad otherwise
\end{array}
\right.
\]
\begin{algorithm}[H]
 \begin{footnotesize}
 \caption{lower layer segment algorithm} \label{alg2}
 \begin{algorithmic}[1]
 \renewcommand{\algorithmicrequire}{\textbf{Input:}}
 \renewcommand{\algorithmicensure}{\textbf{Output:}}
 \REQUIRE $R_{k-1,B}$: Bitrate of the segment in block $k-1$ \\
 		  $\mathcal{B}_k$: Current buffer occupancy \\
 		  $r$ and $c_1$: Sizes of the reservoir and the cushion
 \ENSURE  $R_{k,B}$: Bitrate of the segment in block $k$
  \IF {$R_{k-1,B}=R_{max}$}
  	\STATE $R_{+}=R_{max}$
  \ELSE
  	\STATE $R_{+}=min\{ R_i: R_i > R_{k-1,B}\}$
  \ENDIF
  \IF {$R_{k-1,B}=R_{min}$}
  	\STATE $R_{-}=R_{min}$
  \ELSE
  	\STATE $R_{-}=max\{ R_i: R_i < R_{k-1,B}\}$
  \ENDIF
  \IF {$\mathcal{B}_k \leq r$}
  	\STATE $R_{k,B}=R_{min}$
  \ELSIF {$\mathcal{B}_k \geq r+c_1$}
  	\STATE $R_{k,B}=R_{max}$
  \ELSIF {$F_1(\mathcal{B}_k) \geq R_{+}$}
  	\STATE $R_{k,B}=max\{R_i: R_i < F_1(\mathcal{B}_k)\}$
  \ELSIF {$F_1(\mathcal{B}_k) \leq R_{-}$}
  	\STATE $R_{k,B}=min\{R_i: R_i > F_1(\mathcal{B}_k)\}$
  \ELSE
  	\STATE $R_{k,B}=R_{k-1,B}$
  \ENDIF
  \RETURN $R_{k,B}$
 \end{algorithmic}
  \end{footnotesize}
\end{algorithm}
Our purpose is to increase the video quality and decrease the quality variations. We do this in two steps.

\begin{figure}[t!]
\begin{center}
\includegraphics[scale=0.7]{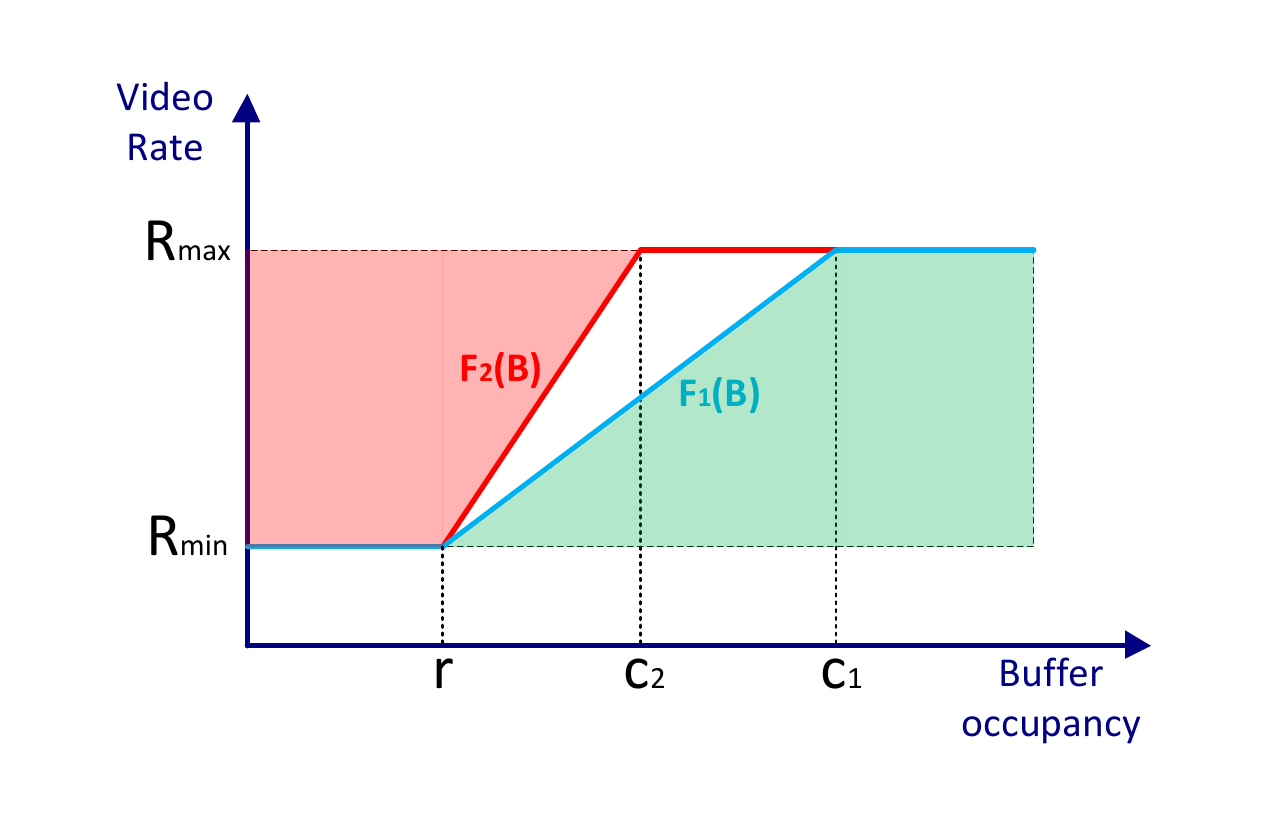}
\caption{The adjusment functions for the lower layer segments and the top layer segments: rates above the curve are risky for buffer depletion, rates below the curve are safer but correspond to lower quality.}
\label{adjusfunc}
\end{center}
\end{figure}

First, we remark that when using BBA-1 algorithm on the lower layer segments in BSC system with the adjustment function $F$, we still have a margin which can be used to add enhancement layers segments while avoiding the playback interruptions. Therefore we define two adjustment functions $F_1$ and $F_2$. The two functions have the same formula as function $F$ but differ in the value of $c$, i.e, $F_1$ uses $c_1$ and $F_2$ uses $c_2$ (Fig. \ref{adjusfunc}). Given the values of $c_1$ and $c_2$, we can increase and decrease the margin between the two curves and then adjust the amount of enhancement layers segments we add to the lower layer ones.

The inputs of the algorithm are the bitrate of the previous lower layer segment, the current buffer occupancy 
and the buffer thresholds. The output is the bitrate of the next lower layer segment. Then, we compute the bounds of the previous bitrate ($R_{+}$ and  $R_{-}$) and the adjustment function $F_1$ regarding the buffer occupancy $\mathcal{B}_k$. The bitrate of the next lower layer segment is selected according to $F_1(\mathcal{B}_k)$ and the buffer occupancy.

\subsection{Smoothing the bitrate variability} 

The main purpose of the enhancement layers segments is to improve the quality of the video. They do not increase the video content in the buffer in terms of playout time. For the first $\phi-1$ blocks, the buffer level increases by two segments after the download of block $k$. Whereas for $k \geq \phi$, the buffer level increases by only one segment after the download of block $k$ is completed. We use the adjustment function $F_2$ to select the bitrate of the enhancement layers segments. Since $F_2 \geq F_1$, we will increase the video quality. But we also want to decrease the quality variations. For this purpose, we will apply the Alg.~\ref{alg3} not on a single enhancement layer segment, but on a set of blocks of enhancement layers segments of length $\phi-1$.

\begin{figure}[t]
\begin{center}
\includegraphics[scale=0.5]{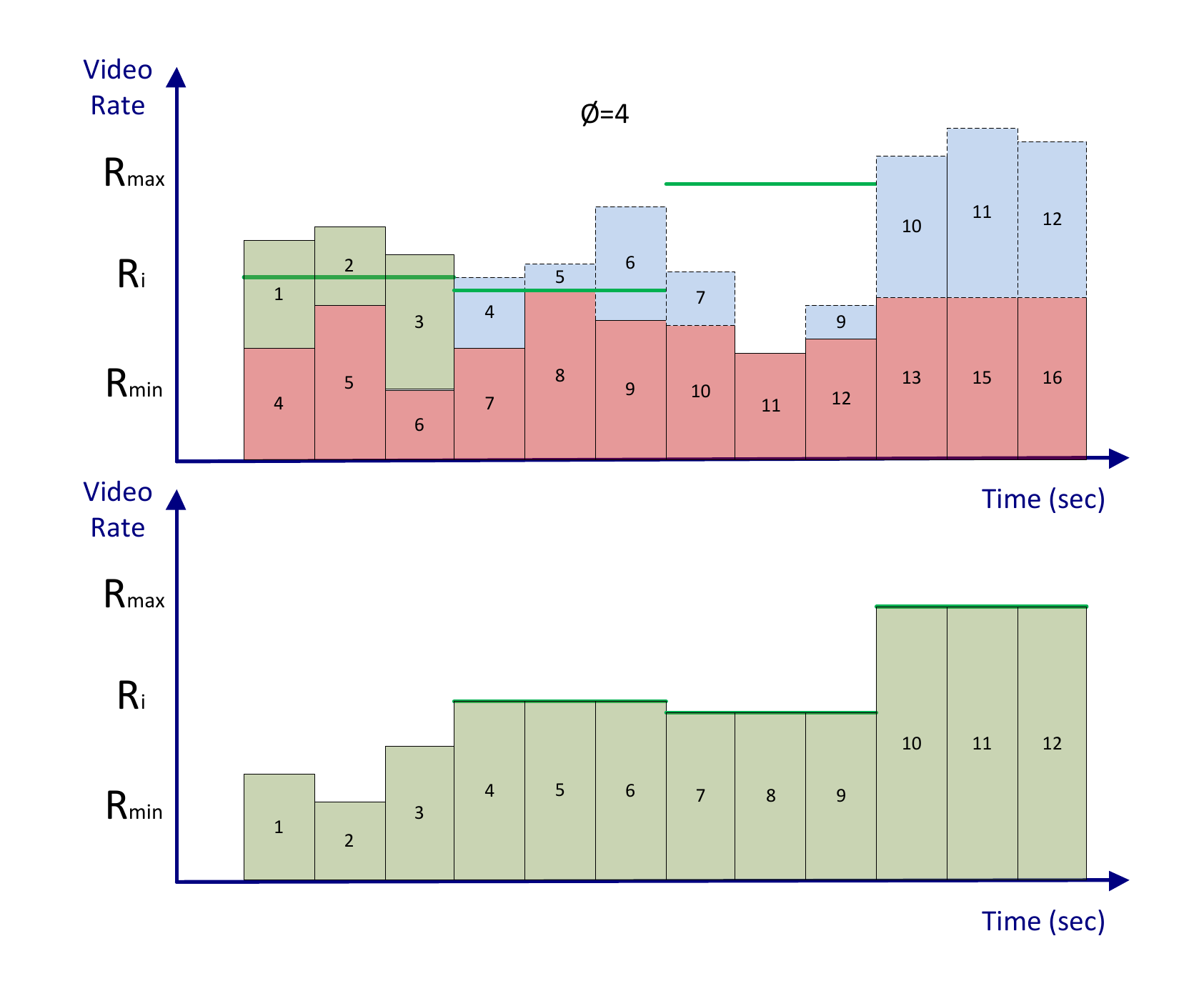}
\put(-238,182){a)}\put(-238,95){b)}
\caption{a) Example of application of Alg.~\ref{alg3}: applied on several segments simultaneously it smooths the bitrate variability b) Effect at the decoder side; $\phi=4$. }
\label{blockalgo}
\end{center}
\end{figure}

An example of this smoothing procedure is reported in Fig. \ref{blockalgo} for $\phi=4$. The algorithm is 
applied on blocks of 3 consecutive enhancement layers segments. The red part represents the lower layer segments. After the download of top layer segment $3$, the output of the algorithm is $R_i$ (the green bar). That means, we have to download the necessary enhancement layers of segment $4$, $5$ and $6$ to reach $R_i$. These enhancement layers will be download on low layer segment $7$, $8$ and $9$ respectively.

\begin{algorithm}[H]
\begin{footnotesize}
 \caption{top layer segment algorithm} \label{alg3}
 \begin{algorithmic}[1]
 \renewcommand{\algorithmicrequire}{\textbf{Input:}}
 \renewcommand{\algorithmicensure}{\textbf{Output:}}
 \REQUIRE $R_{k-1,E}$: Bitrate of enhancement layers segments of the previous $\phi-1$ blocks \\
 		  $\mathcal{B}_k$: Current buffer occupancy \\
 		  $r$ and $c_2$: Sizes of the reservoir and the cushion
 \ENSURE  $R_{k,E}$: Bitrate of enhancement layers segments of the next $\phi-1$ blocks
  \IF {$\mod(k-1,\phi-1)==0$}
  	\IF {$R_{k-1,E}=R_{max}$}
  		\STATE $R_{+}=R_{max}$
  	\ELSE
  		\STATE $R_{+}=min\{ R_i: R_i > R_{k-1,E}\}$
  	\ENDIF
  	\IF {$R_{k-1,E}=R_{min}$}
  		\STATE $R_{-}=R_{min}$
  	\ELSE
  		\STATE $R_{-}=max\{ R_i: R_i < R_{k-1,E}\}$
  	\ENDIF
	\STATE $r_{avg}^{+} \gets [\sum_{i=k-2\phi+2}^{k-\phi}{R_{i,B}+max(0,(R_{+}-R_{i,B}))}]/(\phi-1)$
	\STATE $r_{avg}^{-} \gets [\sum_{i=k-2\phi+2}^{k-\phi}{R_{i,B}+max(0,(R_{-}-R_{i,B}))}]/(\phi-1)$
	\IF {$\mathcal{B}_k \leq r$}
		\STATE $R_{k,E}=R_{k-\phi+1,B}$ //no enhancement layers
	\ELSIF {$\mathcal{B}_k \geq r+c_2$}
		\STATE $R_{k,E}=R_{max}$
	\ELSIF {$F_2(\mathcal{B}_k) \geq r_{avg}^{+}$}
		\STATE $R_{k,E}=max\{R_i: R_i < F_2(\mathcal{B}_k)\}$
	\ELSIF {$F_2(\mathcal{B}_k) \leq r_{avg}^{-}$}
		\STATE $R_{k,E}=min\{R_i: R_i > F_2(\mathcal{B}_k)\}$
	\ELSE
		\STATE $R_{k,E}=R_{k-1,E}$
	\ENDIF
  \ELSE
  	\RETURN $R_{k-1,E}$
  \ENDIF
 \end{algorithmic}
 \end{footnotesize}
\end{algorithm}

The algorithm is invoked after a set of blocks of length $\phi-1$. Then, when the algorithm is invoked after the download of block $k-1$, the output remains the same for the next $\phi-1$ BSC blocks ($\mod(k-1, \phi-1)$). The inputs are the previous enhancement layers segment bitrate, the buffer occupancy and the buffer thresholds $r$ and $c_2$. The output is the bitrate of the next $\phi-1$ enhancement layers segments. For the algorithm of the lower layer segments, we compare $F_1(\mathcal{B}_k)$ to the bounds of the bitrate of the previous lower layer segment. Here, we compare $F_2(\mathcal{B}_k)$ to $r_{avg}^{+}$ and $r_{avg}^{-}$. $r_{avg}^{+}$ ($r_{avg}^{-}$) is the bitrate of each segment in the the previous set of blocks of length $\phi-1$ to reach $R_+$ ($R_-$) where $R_+$ ($R_-$) is the upper (lower) bound of the previous bitrate $R_{k-1,E}$. In other words, $r_{avg}^{+}=R_+$ and $r_{avg}^{-}=R_-$. Then, we compute the bounds of the previous bitrate and the adjustment function $F_2$ corresponding to buffer occupancy $\mathcal{B}_k$. The bitrate of the next enhancement layers segment is selected according to $F_2(\mathcal{B}_k)$ and the buffer occupancy. 

\section{Simulations and Numerical results}
\label{num}
\subsection{Simulation framework}
We evaluate the bitrate adaptation in TB-BSC and BB-BSC using a custom simulation framework. The simulation takes as inputs the network capacity, the number of segments in the video file $N$, the segment duration $L$, the offset $\phi$ and the set of available bitrates $\mathcal{R}$. Then, we model the video downloading, the playback process and the buffer dynamics. The startup delay corresponds to the download time of the first BSC block. After the end of the startup phase, the playback starts with a display speed of $d$ frames per second (fps). We compute the size of each block $k$ according to the bitrates $R_{k,B}$ and $R_{k,E}$. Then the block is downloaded using the real network capacity. At the end of the download of block $k$, we select the bitrates of the next block $k+1$. At the user side, the block is decoded once it is completely downloaded and the segments are available in the buffer.

\begin{figure*}[htb!]
\begin{center}
\begin{minipage}[t]{0.32\linewidth}
\centering\epsfig{figure=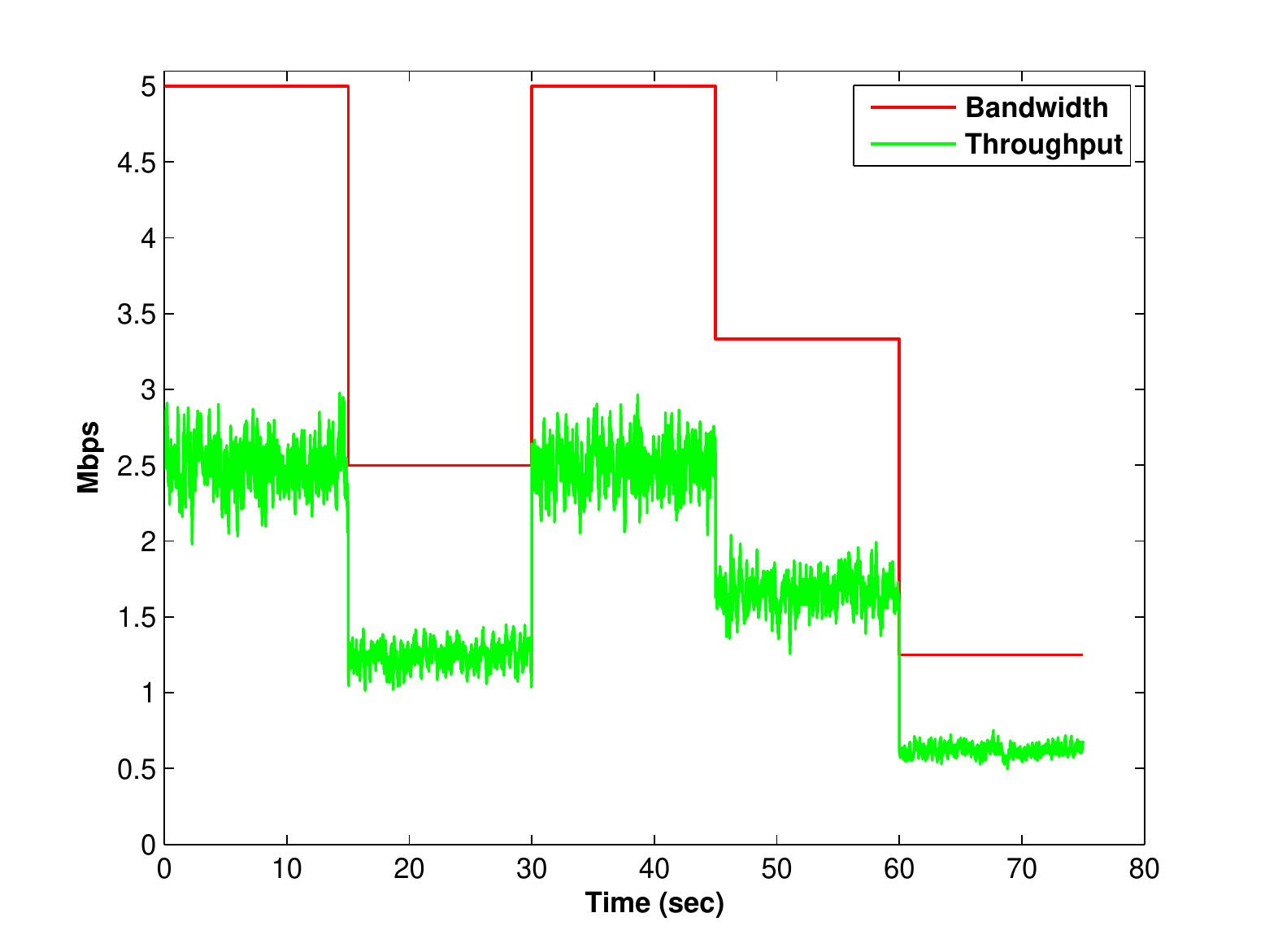, width=\linewidth}
\caption{Background traffic with state variations \label{bandwidth}}
\end{minipage} \hfill
\begin{minipage}[t]{0.32\linewidth}
\centering\epsfig{figure=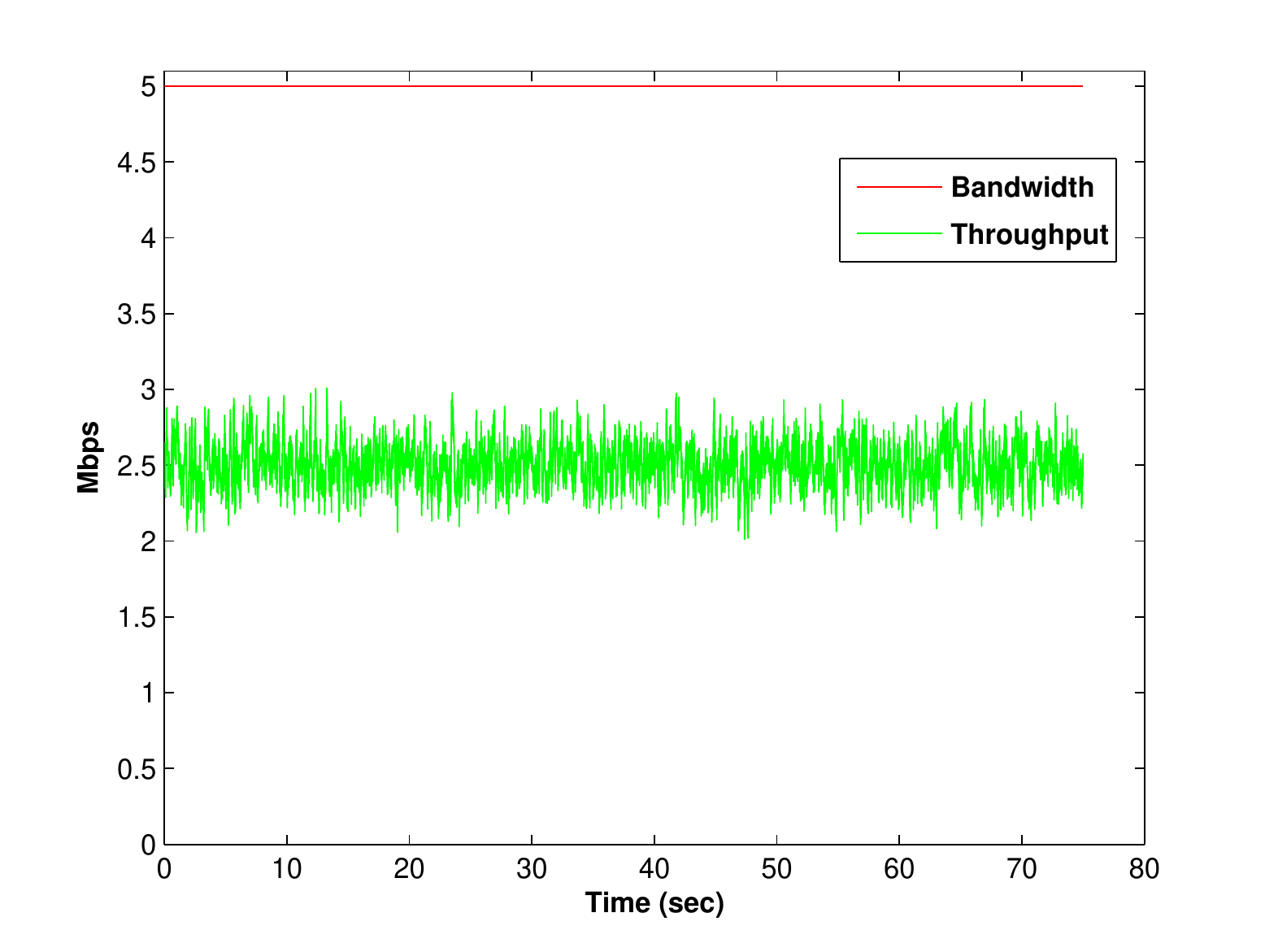, width=\linewidth}
\caption{Background traffic with permanent state \label{bandwidth2}}
\end{minipage} \hfill
\begin{minipage}[t]{0.32\linewidth}
\centering\epsfig{figure=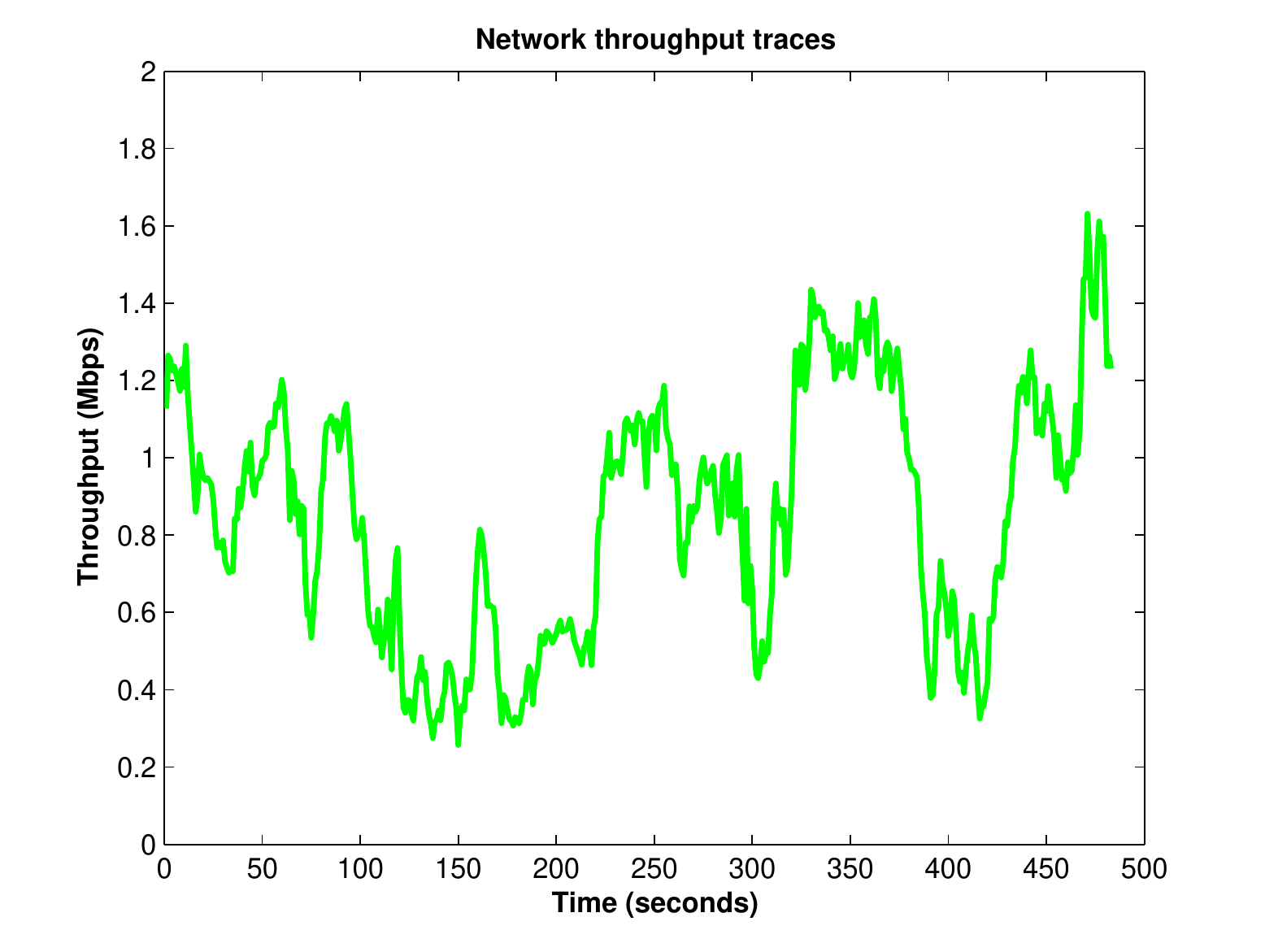, width=\linewidth}
\caption{Network throughput trace from 3G/HSDPA mobile wireless network \label{trace}}
\end{minipage}
\end{center}
\end{figure*}

We test our algorithms under three network capacity variation models. The first two  are a generated network traffic (Fig. \ref{bandwidth} and Fig. \ref{bandwidth2}). Our generated traffic model is similar to that of \cite{tian2012towards} which is a realistic network bandwidth variations with different congestion states, 
where background TCP traffic is injected between the server and the client. We set the link capacity between 
the server and the client to be 5Mbps. For the generated network traffic of Fig.~\ref{bandwidth}, we simulate 
the behaviour of a standard linux \textit{traffic control tool} by decreasing or increasing the throughput. The traffic rate thus jumps between different states. The oscillations within the same state are due to TCP congestion control mechanisms. 
The link capacity is higher than the TCP throughput: we assume that this throughput is used only for the video segments transmission since the audio segments and the synchronisation data can still use a fraction of the whole link capacity. 

The third traffic model is a real HSDPA trace reporting logs from TCP streaming sessions in Telenor's 3G/HSDPA mobile wireless network in Norway \cite{hsdpa_trace}(Fig. \ref{trace}). It is an average throughput trace collected from several measurements inside a tramway from Ljabru (start location) to Jernbanetorget (destination location). For each measurement, adaptive video streams were downloaded at the maximum speed with a video chunk duration of 2 seconds.
 
\subsection{Numerical Results}
\begin{figure*}[htb!]
\begin{center}
\begin{minipage}[t]{0.4\linewidth}
\centering\epsfig{figure=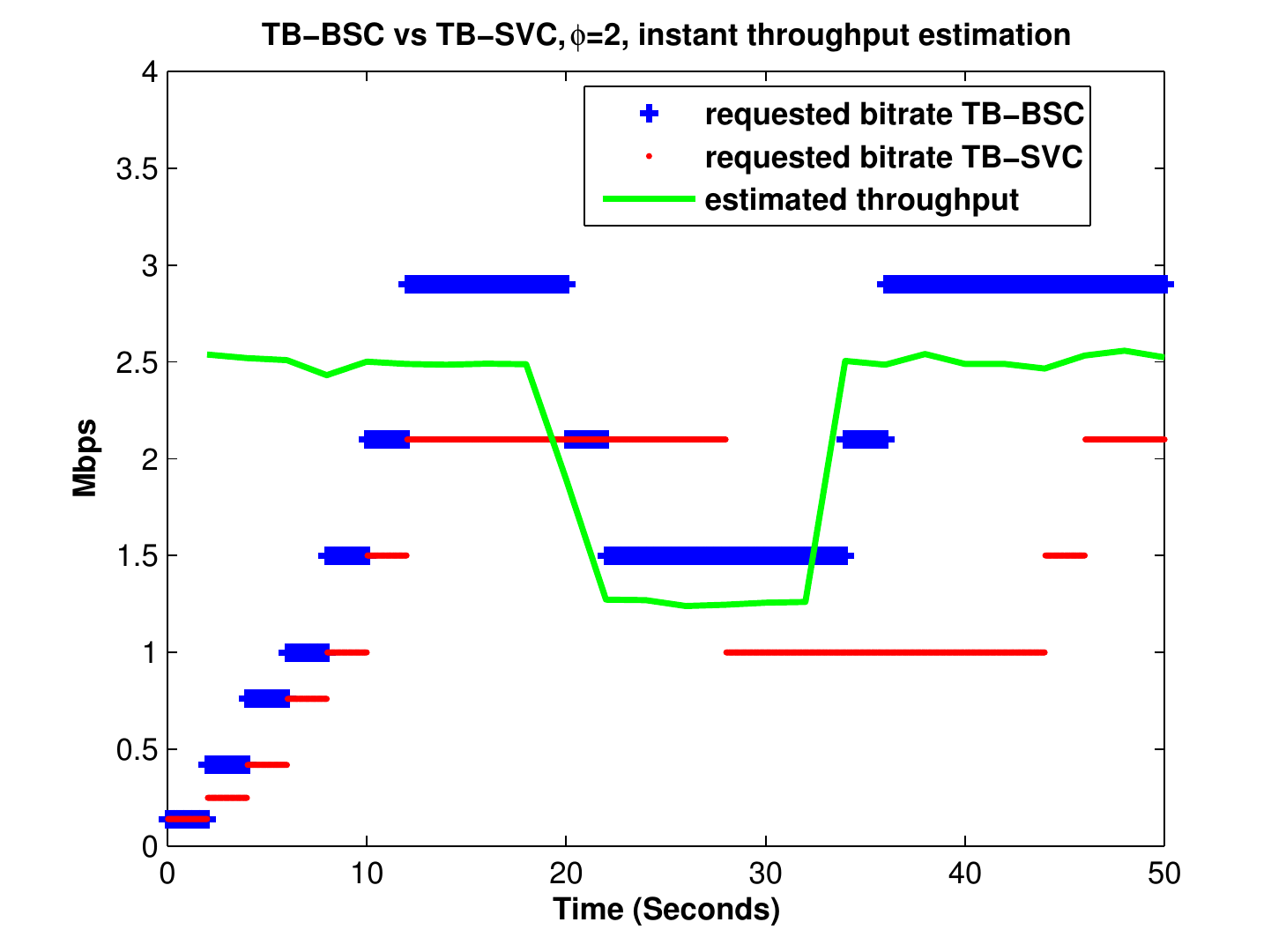, width=\linewidth}
\caption{Requested bitrates for TB-BSC and TB-SVC for instant throughput estimation method \label{instant}}
\end{minipage} \hfill
\begin{minipage}[t]{0.4\linewidth}
\centering\epsfig{figure=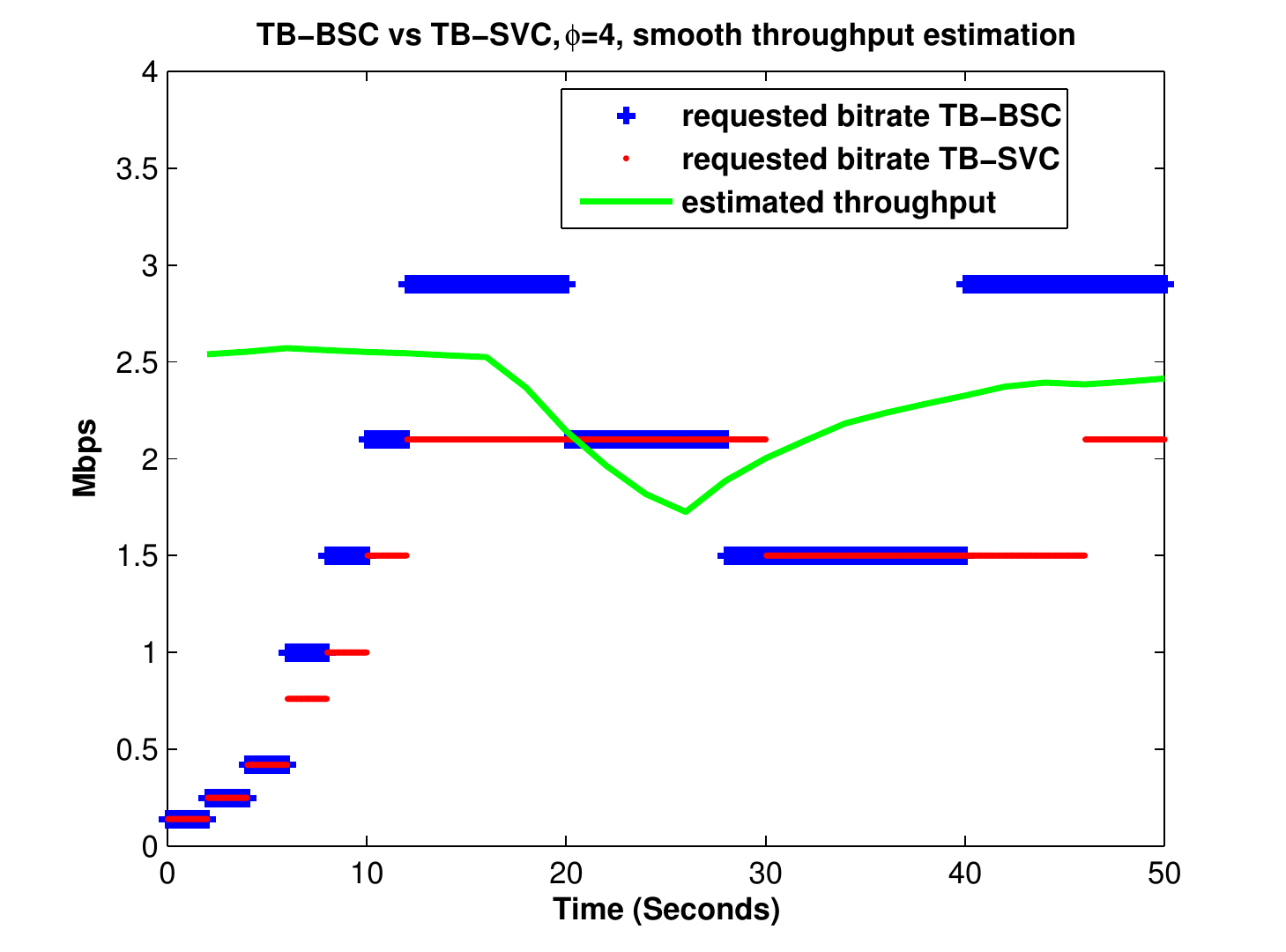, width=\linewidth}
\caption{Requested bitrates for TB-BSC and TB-SVC for smooth throughput estimation method \label{stable}}
\end{minipage} 
\hfill
\begin{minipage}[t]{0.4\linewidth}
\centering\epsfig{figure=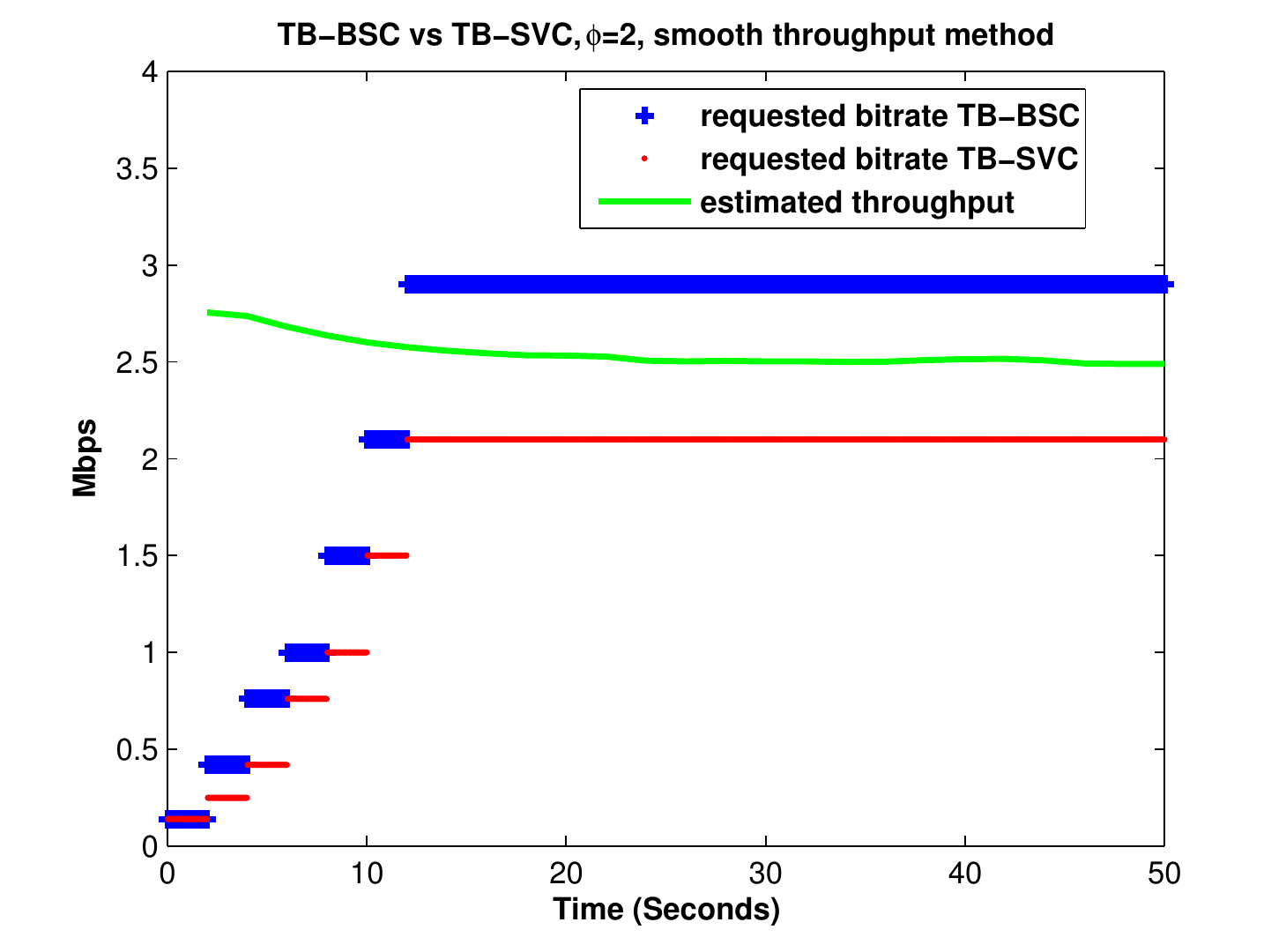, width=\linewidth}
\caption{Requested bitrates for TB-BSC and TB-SVC with permanent bandwidth state \label{permanent}}
\end{minipage} \hfill
\begin{minipage}[t]{0.4\linewidth}
\centering\epsfig{figure=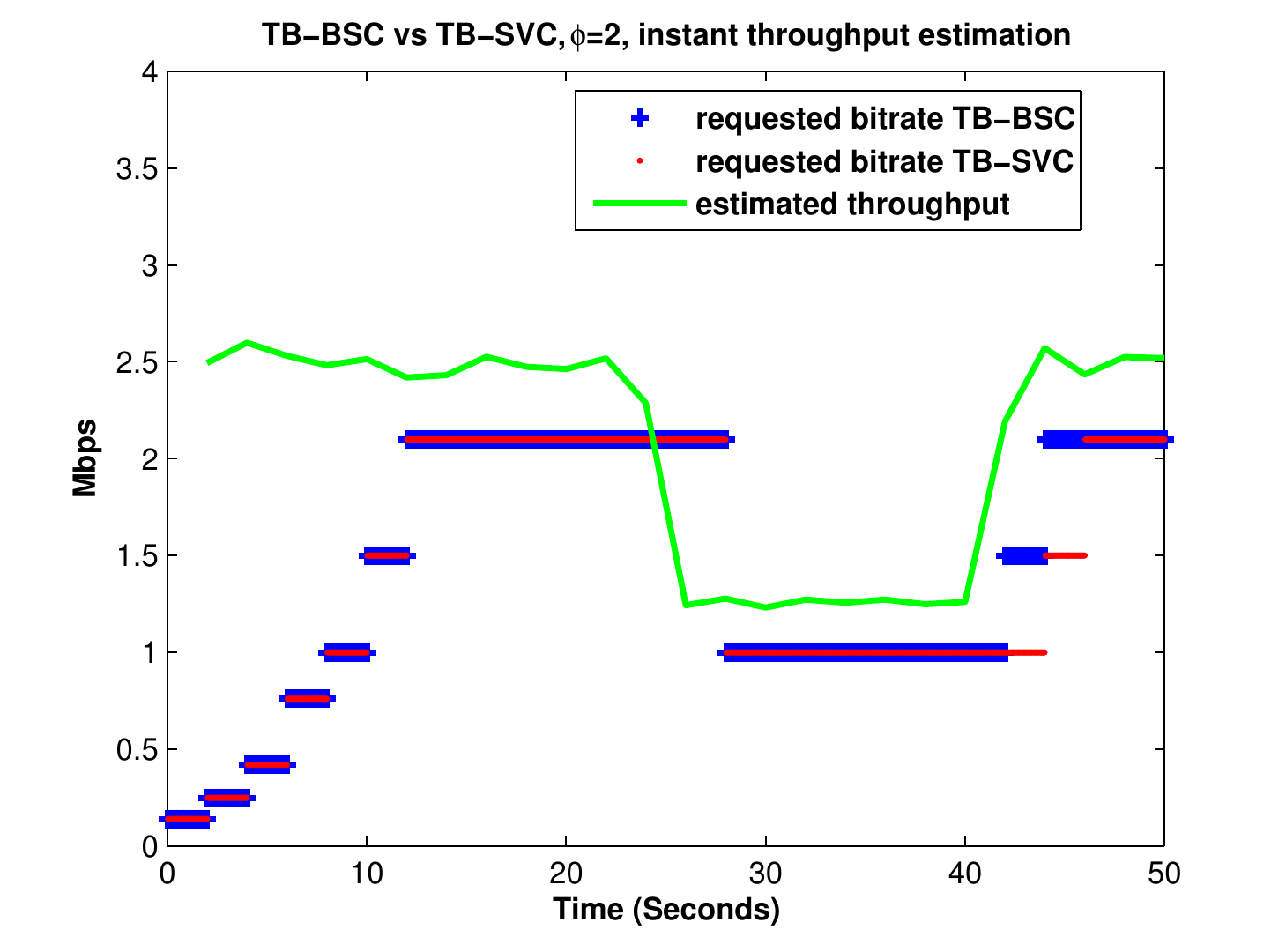, width=\linewidth}
\caption{Requested bitrates for TB-BSC and TB-SVC \label{conser}}
\end{minipage}
\end{center}
\end{figure*}
The set of experiments compare the requested bitrate with TB-BSC  and  the throughput based scheme using SVC  (TB-SVC) under the network conditions of Fig. \ref{bandwidth}, Fig. \ref{bandwidth2} and the throughput traces of Fig. \ref{trace}. The file size in the experiments is up to 350 seconds ($>$ 5 minutes) of video while the playback frequency is 25 frames per second (fps). We consider the following set of available bitrates $\{$ 140, 250, 420, 760, 1000, 1500, 2100, 2900 $\}$ (Kbps). The video segment duration is set to $2$ seconds.

Fig. \ref{instant} shows the requested bitrate for TB-BSC and TB-SVC for the throughput based algorithm where the throughput is estimated over the last segment that was downloaded (instant throughput). TB-BSC achieves a higher average bitrate compared to TB-SVC for only one additional quality variations. The average bitrate is $2040$ Kbps with TB-BSC and $1382$ Kbps with TB-SVC. As shown in the Fig. \ref{instant}, we begin with the smallest bitrate and we increase the bitrate in a smooth manner because according to \cite{qdash}, users prefer a gradual quality change. The number of quality variations is $10$ for TB-BSC and $9$ for TB-SVC.

Fig. \ref{stable} shows the requested bitrate for TB-BSC and TB-SVC for the throughput based algorithm where the throughput is estimated over all the previous segments (smooth throughput). We use the estimation method of \cite{akhshabi2011experimental} where the actual estimated throughput is 20\% of the throughput estimated over the last segment and 80\% of the estimated throughput over all the previous segments. The purpose of this method of estimation is to cope with the throughput short-term fluctuations. As for the instant throughput case, TB-BSC system achieves a better video quality than TB-SVC. The average bitrate is $1948$ Kbps for TB-BSC and $1566$ Kbps for TB-SVC. The number of quality variations is $9$ for TB-BSC and $8$ for TB-SVC. If we compare these results with the instant throughput case, we see that the number of quality variations decreases. This method of throughput estimation is more robust than the instant throughput case.

For the permanent bandwidth state case (the bandwidth stays at 5Mbps in Fig. \ref{bandwidth2}), the throughput is around 2.5Mbps during the entire video session. Recall that the small throughput fluctuations are due to TCP congestion control mechanisms. We compute the requested bitrate in TB-BSC and TB-SVC in that case. Fig. \ref{permanent} shows the results. The first observation is that BSC definitely outperforms classical DASH system since the average bitrate is $2440$ Kbps for BSC against $1758$ Kbps for DASH. Moreover, the two systems have the same number of quality variations (e.g., $6$). The estimated throughput is around $2.5$ Mbps. For the same network conditions, TB-BSC renders the video with $2900$ Kbps quality while TB-SVC cannot. This is the main insight of the Backward-Shifted Coding system: \textit{render high quality video under limited network bandwidth}. If we force TB-SVC to play the video with the quality of $2900$ Kbps, we will have a severe number of the playback interruptions. 

 Table \ref{tabgen} gives the average quality, the variance of the quality, the number of quality variations, and the number of the playback interruptions for $100$ simulations for three scenarios: {\bf scenario 1} is the permanent bandwidth state case with the smooth throughput estimation method while {\bf scenario 2}  and {\bf scenario 3}  are, respectively, the variable bandwidth state case with the instant and the smooth throughput estimation methods.
\begin{table}[h]
\begin{center}
\begin{tabular}{|p{1.5cm}|p{1cm}|p{1cm}|p{1cm}|p{1.2cm}|}
\hline
& \scriptsize Average quality (Kbps) & \scriptsize Variance of the quality & \scriptsize Number of switchings & \scriptsize Number of playback interruptions \\
\hline
\scriptsize Scenario 1 TB-BSC & \scriptsize \textbf{2324} & \scriptsize 825e9 & \scriptsize 11.84 & \scriptsize \textbf{0} \\
\scriptsize Scenario 1 TB-SVC & \scriptsize 1758 & \scriptsize \textbf{421e9} & \scriptsize \textbf{6} & \scriptsize \textbf{0} \\
\hline
\scriptsize Scenario 2 TB-BSC & \scriptsize \textbf{1514} & \scriptsize 452e9 & \scriptsize \textbf{9} & \scriptsize \textbf{0} \\
\scriptsize Scenario 2 TB-SVC & \scriptsize 1382 & \scriptsize \textbf{425e9} & \scriptsize \textbf{9} & \scriptsize \textbf{0} \\
\hline
\scriptsize Scenario 3 TB-BSC & \scriptsize \textbf{1951} & \scriptsize 737e9 & \scriptsize 9.73 & \scriptsize 0.29 \\
\scriptsize Scenario 3  TB-SVC& \scriptsize 1567 & \scriptsize \textbf{369e9} & \scriptsize \textbf{8} & \scriptsize \textbf{0} \\
\hline
\end{tabular}
\end{center}
\caption{The average statistics for $100$ simulations for $\phi=4$} \label{tabgen}
\end{table} 
The Backward-Shifted Coding achieves a higher quality but we still have a high  quality variations.

The Backward-Shifted Coding system can behave exactly like TB-SVC system if we choose the same bitrate for the low layer segments $R_{k,B}$ and the enhancement layers segments $R_{k,E}$. Fig. \ref{conser} shows the requested bitrate for TB-BSC and TB-SVC when $R_{k,B}=R_{k,E}$. The requested bitrate is almost the same for the two systems (TB-BSC still has a higher bitrate). This short difference is due to the offset $\phi$ in BSC. This property adds more flexibility to BSC system since one can switch from TB-BSC to TB-SVC or inversely depending on the network capacity.

\begin{figure*}[htb!]
\begin{center}
\begin{minipage}[t]{0.4\linewidth}
\centering\epsfig{figure=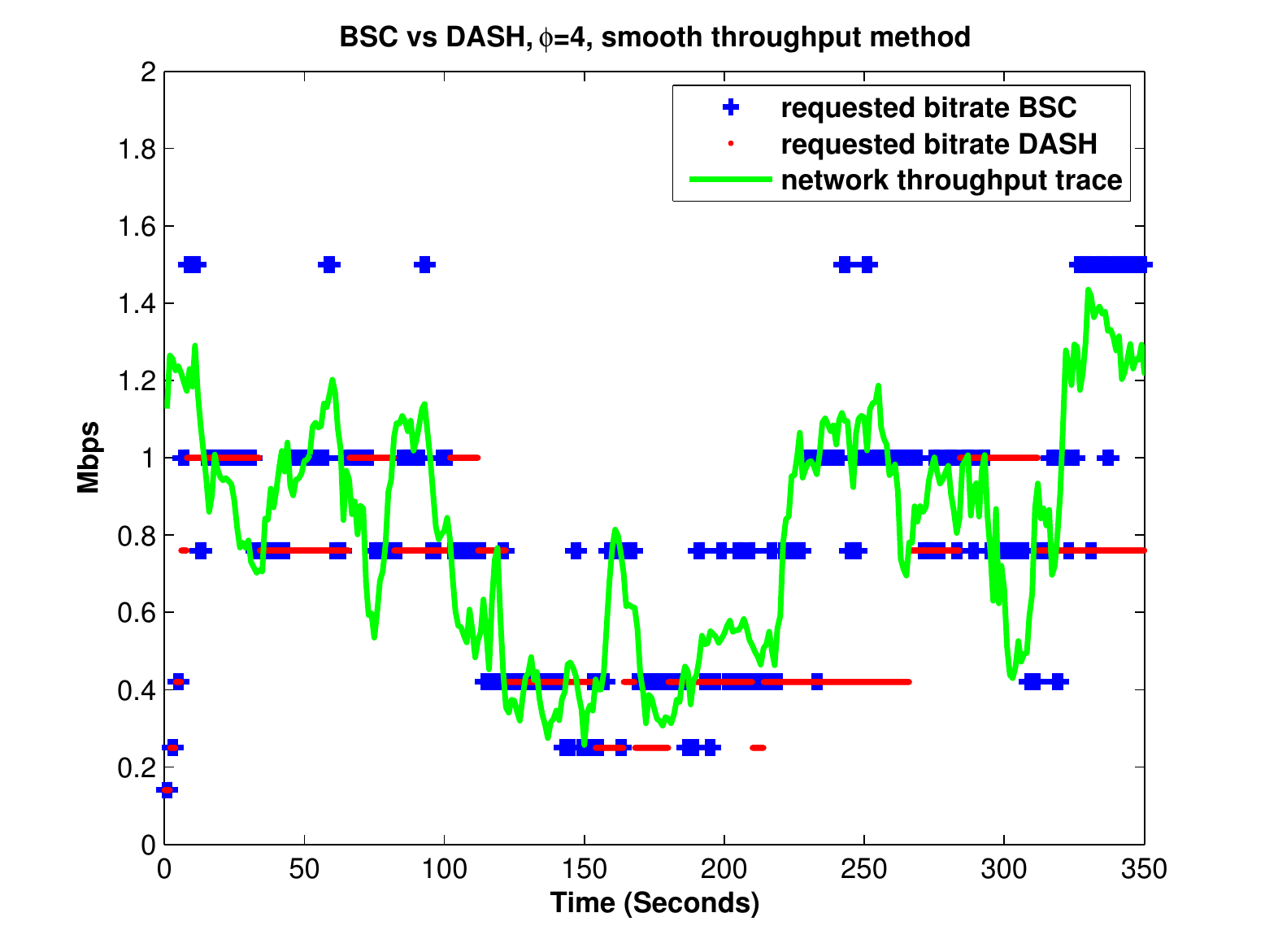, width=\linewidth}
\caption{Requested bitrates for TB-BSC and TB-SVC for smooth throughput estimation method \label{stable_trace}}
\end{minipage} \hfill
\begin{minipage}[t]{0.4\linewidth}
\centering\epsfig{figure=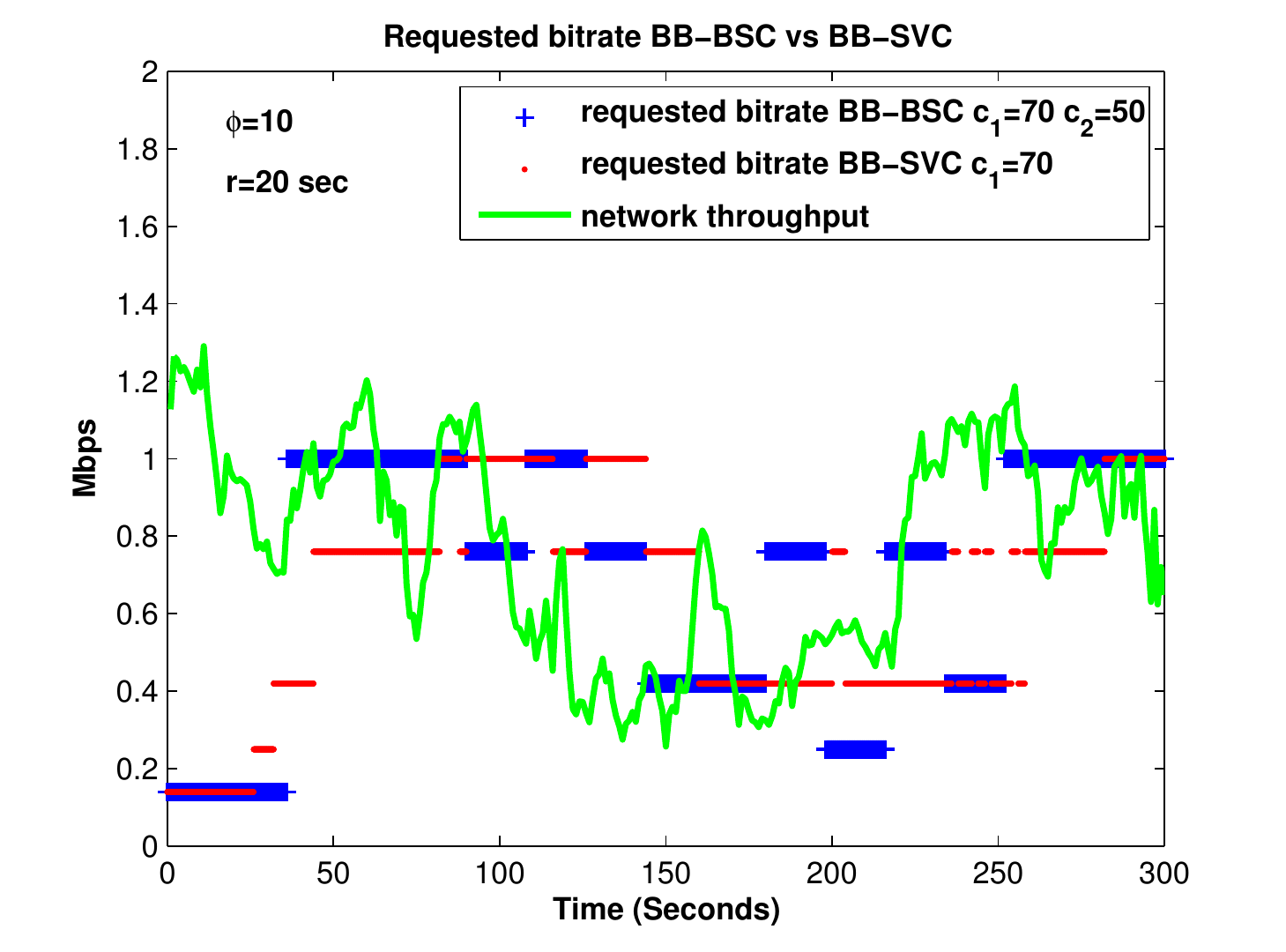, width=\linewidth}
\caption{Requested bitrates for BB-BSC and BB-SVC for buffer based method \label{dashbsc1}}
\end{minipage} 
\hfill
\begin{minipage}[t]{0.4\linewidth}
\centering\epsfig{figure=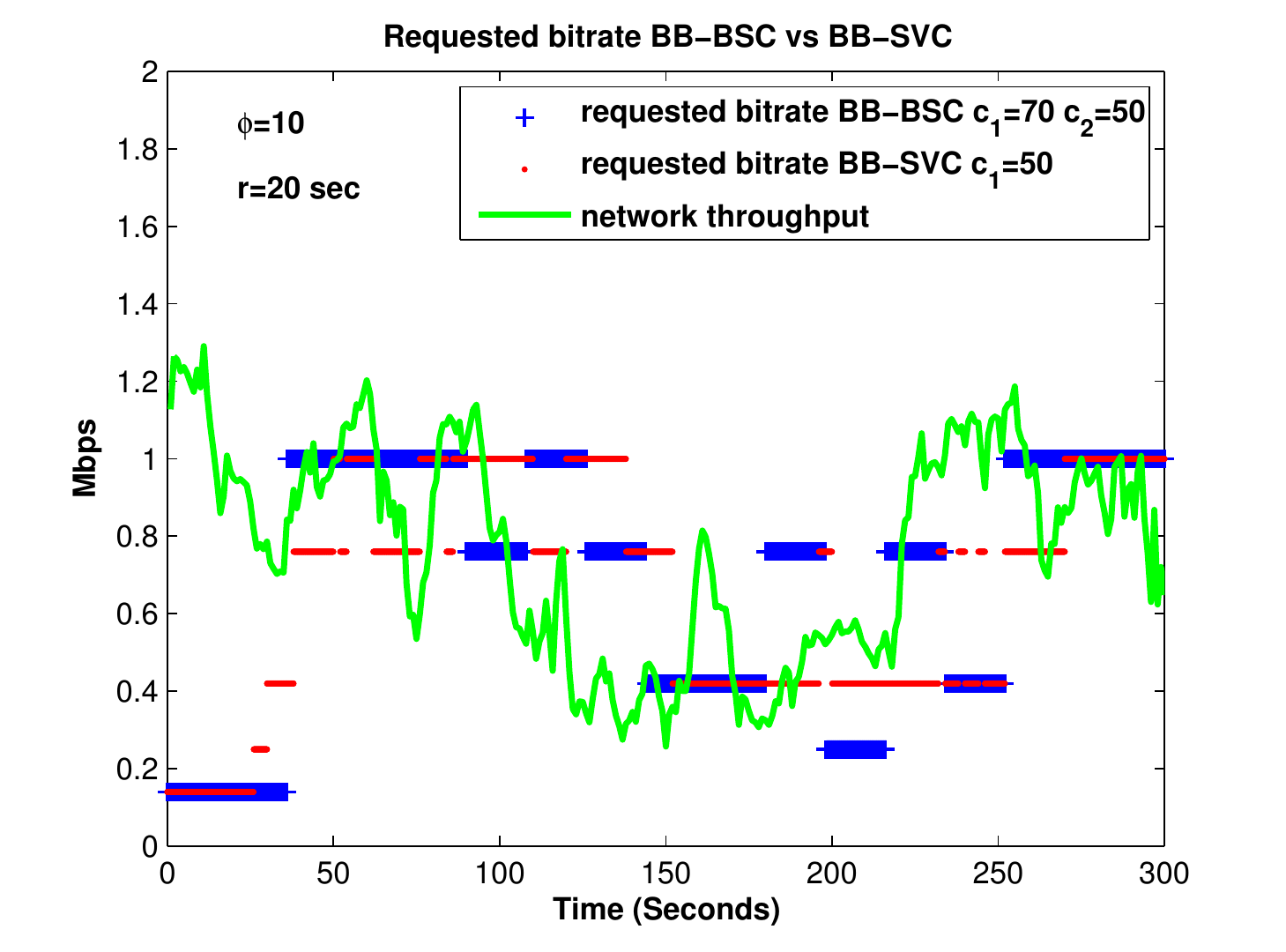, width=\linewidth}
\caption{Requested bitrates for BB-BSC and BB-SVC for buffer based method \label{dashbsc2}}
\end{minipage} \hfill
\begin{minipage}[t]{0.4\linewidth}
\centering\epsfig{figure=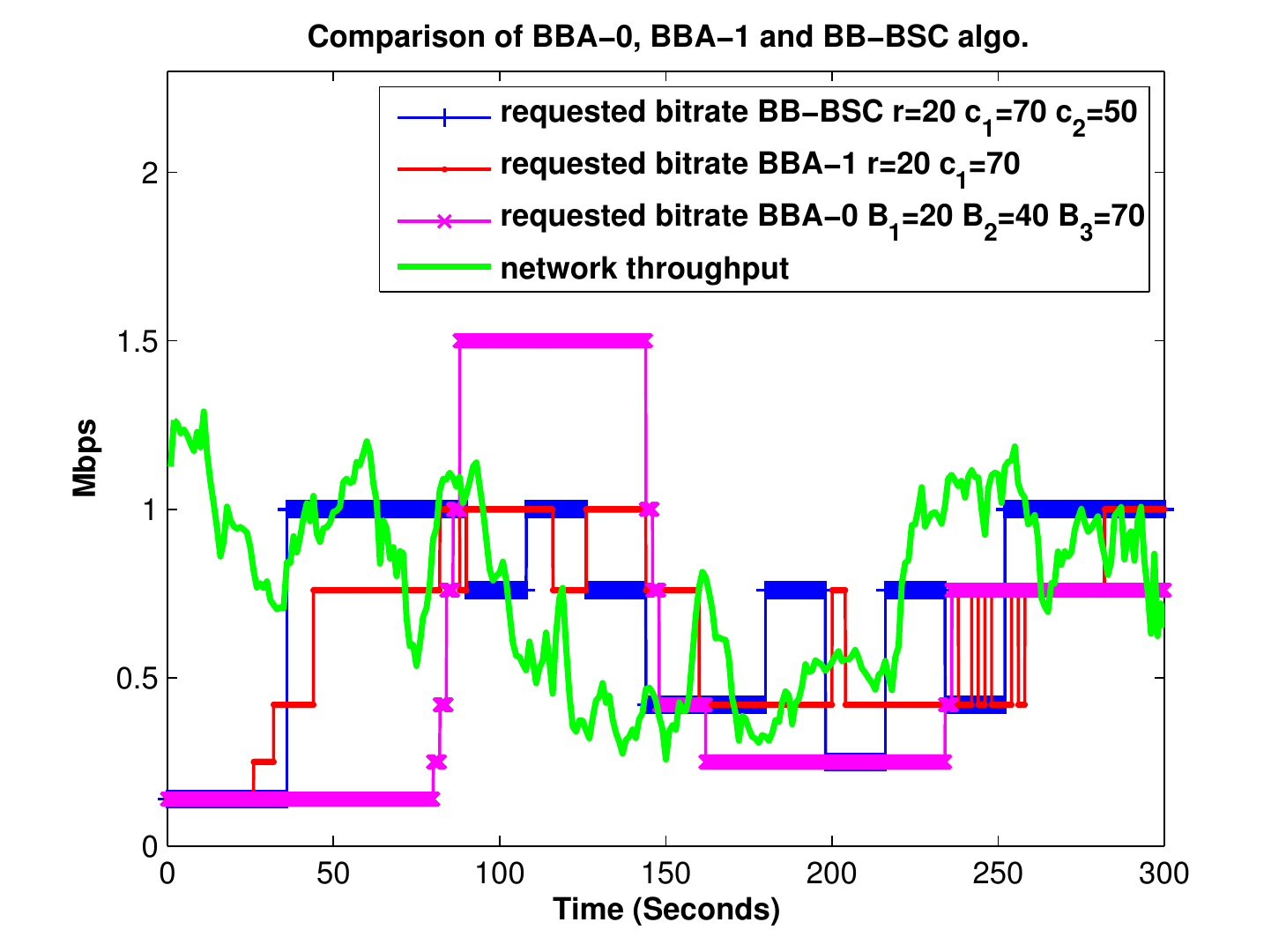, width=\linewidth}
\caption{Comparison of BBA-0, BBA-1 and BB-BSC algorithm \label{comparison}}
\end{minipage}
\end{center}
\end{figure*}
For the 3G/HSDPA network bandwidth case, Fig. \ref{stable_trace} confirms that TB-BSC improves the video quality than TB-SVC. But we still have too much quality variations. The main reason for this is the possible capacity estimation errors due to the short-term capacity fluctuations. The throughput based method has its limitations and we resort to the buffer based method to smooth the quality.

Fig. \ref{dashbsc1} shows the requested bitrate for BB-BSC and BB-SVC for the buffer based method. The buffer thresholds are $r=20$ sec, $c_1=70$ sec, $c_2=50$ sec and the offset $\phi=10$. DASH/BSC achieves a higher video quality than BB-SVC. The average bitrate is $689$ Kbps for BB-BSC against $660$ Kbps for BB-SVC. There is also less quality variations for BB-BSC, $10$ against $22$ for BB-SVC. The BB-SVC algorithm herein is BBA-1. It uses $r$, $c_1$. In the previous figure, $c_1=70$. What happens if $c_1=50$? Fig. \ref{dashbsc2} shows the requested bitrates. The results are similar to the case $c_1=70$, then BB-BSC still outperforms BB-SVC with $10$ quality variations against $22$ for BB-SVC. Fig. \ref{comparison} shows the requested bitrates for BB-SVC, BBA-1 and BBA-0. BBA-0 is the algorithm of \cite{miller2012adaptation}. There are three buffer thresholds $B_1$, $B_2$ and $B_3$. To make the comparison fair, we take $B_1=r$, $B_3=c_1$ and make simulations to select the value of $B_2$ that gives the best results for BBA-0. The results of the comparison are shown in table \ref{tabtrace}.
\begin{table}[h]
\begin{center}
\begin{tabular}{|p{1.5cm}|p{1.2cm}|p{1.4cm}|p{1.4cm}|p{1.5cm}|}
\hline
& Average quality (Kbps) & \scriptsize Variance of the quality & \scriptsize Number of switchings & \scriptsize Number of playback interruptions \\
\hline
\scriptsize BBA-0 & \scriptsize 581 & \scriptsize 281e9 & \scriptsize 13.61 & \scriptsize \textbf{0} \\
\hline
\scriptsize BBA-1 & \scriptsize 634 & \scriptsize 809e9 & \scriptsize 21.38 & \scriptsize \textbf{0} \\
\hline
\scriptsize TB-BSC  & \scriptsize \textbf{698} & \scriptsize \textbf{950e9} & \scriptsize 58 & \scriptsize 0 \\
\hline
\scriptsize BB-BSC & \scriptsize \textbf{687} & \scriptsize \textbf{116e9} & \scriptsize \textbf{10} & \scriptsize \textbf{0} \\
\hline
\end{tabular}
\end{center}
\caption{Average of QoE metrics: Average quality, quality variability, number of switches and number of playback interruption. } \label{tabtrace}
\end{table} 

\subsection{Impact of the Offset $\phi$}
\begin{figure}[hbt!]
\begin{center}
\includegraphics[scale=0.5]{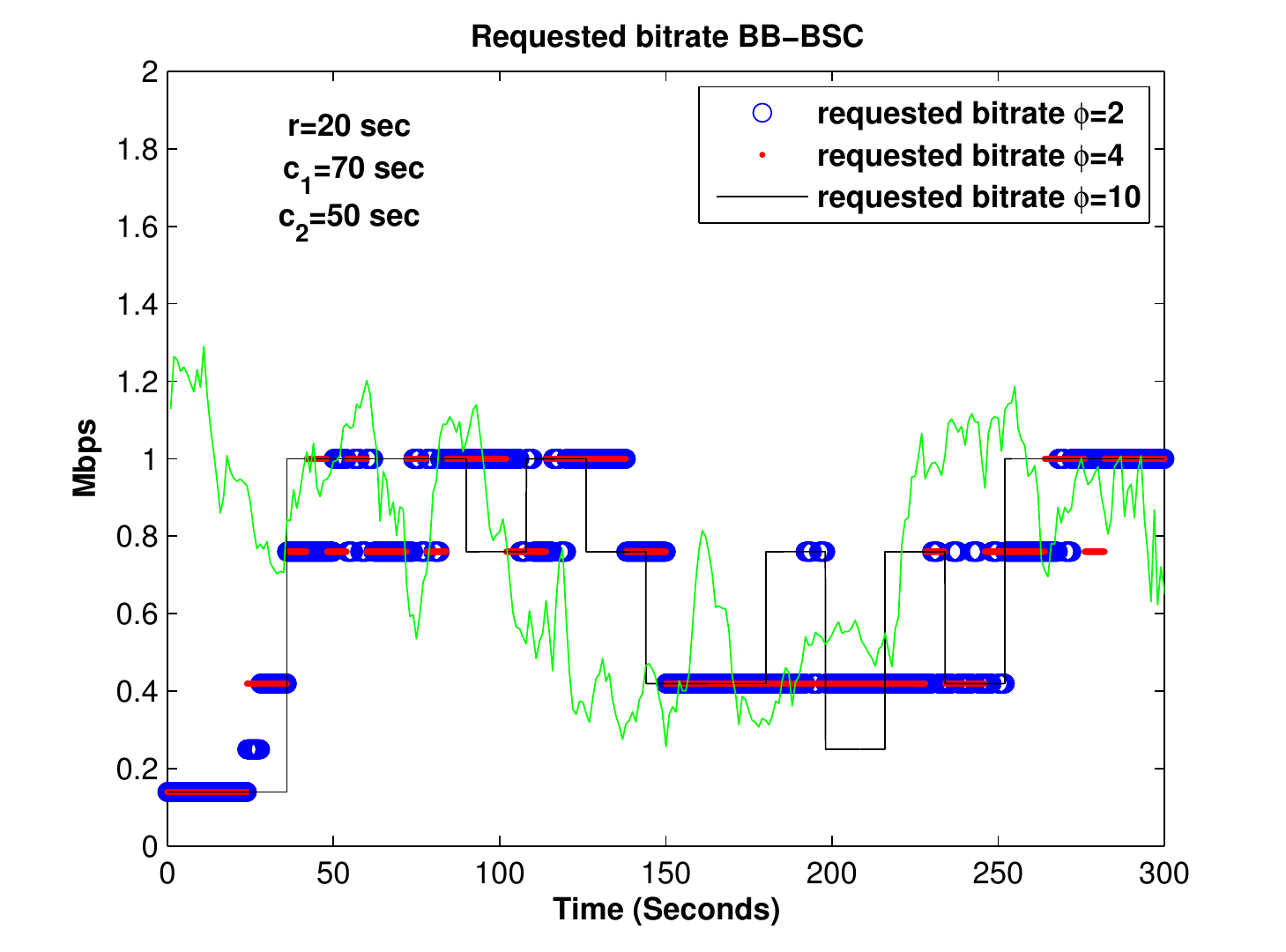}
\caption{Requested bitrates for BB-BSC for different values of $\phi$}
\label{phiimpact}
\end{center}
\end{figure}
The offset $\phi$ has an important role in the Backward-Shifted Coding system. As shown in Fig. \ref{phiimpact}, when $\phi$ increases, the video quality increases as well. The average quality is $662$ Kbps, $666$ Kbps and $689$ Kbps respectively for $\phi=2$, $\phi=4$ and $\phi=10$. Moreover, the number of quality variations decreases. The number of quality variations is $38$, $19$ and $10$ respectively for $\phi=2$, $\phi=4$ and $\phi=10$. One important observation in Fig. \ref{phiimpact} is the first quality variation. When $\phi$ increases, the quality increases quickly at the beginning (the first quality variation). Why? Remember that we say that for $\phi-1$ first BSC blocks, the buffer size increases by two segments after the download of each block. But at that moment it is not yet possible to play the low layer segments since we get segment $\phi-1$. Once we get segment $\phi-1$, the buffer size increases by $\phi$ segments ($\phi L$ seconds of video content). Then, the bitrate jumps quickly also.

\section{Related works}
\label{relatedsec}
Backward-Shifted Coding (BSC) \cite{BSCcoding} and \textit{Streamloading} \cite{streamloading} are two 
schemes which leverage on time redundancy mechanisms for scalable video coding. In both schemes, base 
layer or enhancement layers of any number of future segments of a video can be delivered in advance. 

The analysis reported in~\cite{BSCcoding} provides a complete characterization of BSC's performance
with respect to a set of key metrics, namely the initial buffering delay, the playback interruption, 
and the mean video quality, which are responsible for the users' quality of experience (QoE). Also, 
a cost function is proposed in order to evaluate the QoE. However, the analysis proposed 
in~\cite{BSCcoding} does not account for the concurrent effect of bitrate adaptation. 
  
The \textit{Streamloading} scheme, combines video streaming and video downloading. Users are able to 
download optional enhancement layers for a tagged video. But, video streaming leverages on the base 
layer only. In~\cite{streamloading}, numerical simulations demonstrate that~\textit{streamloading} 
system renders higher average video data rate than standard streaming schemes (HAS); the effect of 
adaptation part is neglected. 

In general, how to choose the bitrate per video layer is an actively investigated research task.
Recent mechanisms have been addressing bitrate adaptation in order to enhance the performance of DASH. 
Thus, in \cite{zou2015can}, authors demonstrate that by knowing the available bandwidth for a few 
seconds in the future, it is possible to improve the bitrate selection. On the same rationale \cite{mekki2015anticipatory}, authors developed an anticipatory HTTP adaptive streaming policy to 
select the bitrate based on a prediction of the wireless channel state. The forecast is  based the 
Received Signal Strength (RSS) of the mobile device. The bitrate is reduced when the RSS decreases 
significantly in order to prevent playback interruptions. 
Adaptation based on throughput estimation -- e.g., by means of direct throughput -- suffers from significant biases since it tends to fluctuate due to the channel short term variations~\cite{thang2014evaluation, zou2015can}. Some approaches try to work around these biases by either smoothing out throughput estimates~\cite{akhshabi2011experimental} or by designing better scheduling strategies \cite{joseph2014nova, khan2015network}. 
 
Adaptation based on buffer occupancy aims at keeping the buffer occupancy at a desired level  \cite{huang2014buffer, julurisara, miller2012adaptation, sobhani2015fuzzy}. In \cite{huang2014buffer}, 
buffer occupancy is showed a more reliable control parameter compared to end-to-end throughput. The 
approach is showed to reduce the rebuffering rate by 20\% and yet deliver higher video quality. In \cite{miller2012adaptation}, the buffer is divided into several occupation ranges. Hence, the goal of the algorithm is to keep the buffer size between two desired thresholds by controlling the bitrate. Authors in \cite{julurisara} use a harmonic filter to estimate the network capacity and then design a controller to drive the buffer size to a reference value. 

In \cite{sobhani2015fuzzy}, both the estimated throughput and the buffer occupancy are used to select the segment bitrate. A set of rules are defined: based on such rules, the bitrate is increased, decreased or kept unchanged. \cite{yin2014toward, yin2015control} develop a model predictive control algorithm to optimally combine throughput and buffer based adaptation methods; selection inaccuracy is ascribed to the throughput estimation errors. 

The bitrate adaptation proposed in this work for BSC differs from those papers in that the adaption policy controls the qualities of two shifted segments at the same time.

\section{Conclusion and Discussion}
\label{conclusion}
In this paper, we studied a novel bitrate adaptation for the Backward-Shifted Coding (BSC) scheme proposed in \cite{BSCcoding}. The Backward-Shifted Coding is based on scalable video coding. The main idea is to split the segments into low layer segments and top layers segments, and send the low layer segments in advance. Depending on the network capacity, the quality of these segments can be improved later by sending only the appropriate number of enhancement layers. In general, the main challenge for scalable video coding schemes is bitrate adaptation, i.e., how to match the quality of the base layer segments and the enhancement layers segments to variable network conditions. We have proposed two bitrate adaptation algorithms, namely TB-BSC and BB-BSC, which have been designed on top of BSC. They are based on network throughput measurements and playback buffer occupancy level, respectively. We further performed simulations compare the efficiency of BSC adaptation methods  to existing DASH/SVC solutions. The results show that our BSC adaptation methods achieve better video quality under  same network conditions, thus providing a DASH-compliant solution rendering high quality video in HTTP adaptive streaming at low cost.


\bibliographystyle{ieeetr}
\bibliography{ZakReferences}

\end{document}